\documentstyle[pra,aps]{revtex} 
\begin{document} 
  \draft 
  \title{IS COSMIC COINCIDENCE A CONSEQUENCE OF A LAW OF NATURE?}
\author
{E. I. Guendelman \thanks{guendel@bgumail.bgu.ac.il} and
A.  B.  Kaganovich \thanks{alexk@bgumail.bgu.ac.il}} 
\address{Physics Department, Ben Gurion University of the Negev, Beer
Sheva 
84105, Israel}
\maketitle
\begin{abstract} 
  A field theory is proposed where the regular fermionic matter and the dark
fermionic matter are different states of the same "primordial" fermion fields.
In regime of the fermion densities typical for normal particle physics, the
primordial fermions split into three families identified with regular fermions.
When fermion energy density becomes comparable with dark energy density, the theory
allows new type of states. The possibility of such Cosmo-Low Energy Physics
(CLEP) states is demonstrated by means of solutions of the field theory
equations describing FRW universe filled by homogeneous scalar field and 
uniformly distributed nonrelativistic neutrinos. Neutrinos in CLEP state
are drawn into
cosmological expansion by means of dynamically changing their own parameters.
One of the features of the fermions in CLEP state is that in the late time
universe their masses
increase as $a^{3/2}$ ($a=a(t)$ is the scale factor). The energy
density of the cold dark matter consisting of neutrinos in CLEP state scales as
a sort of dark energy; this cold dark matter possesses negative pressure and for the
late time universe its equation of state approaches that of the cosmological
constant. The total energy density of such universe is less than it would be in
the universe free of fermionic matter at all. The (quintessence) scalar field
is coupled to dark matter but its coupling to regular fermionic matter 
appears to be
extremely strongly suppressed. The key role in obtaining these results
belongs  to a fundamental
constraint (which is consequence of the action principle) that plays the role
of a new law of nature.

   \end{abstract}          
    
    \renewcommand{\baselinestretch}{1.6} 

\pagebreak 

\section{Introduction} 

Results of cosmological observations\cite{Riess},\cite{Spergel} strengthen  
confidence
 in the truth of a model according to which our universe is spatially flat  and 
composed by $2/3$ of  the so-called dark energy\cite{de-review}
(responsible for the 
present accelerated expansion)
and by $1/3$ of the cold dark matter\cite{dm-review}
 responsible for the flat rotation curves of
galaxies and  for structure formation at large scales.
  The notion of dark energy is usually used as a label for either 
positive
cosmological constant or  a scalar field slowly evolving down a potential
(quintessence)\cite{quint}.
The attribute of dark energy is negative pressure.  
 In contrast with this,
  cold dark matter is expected naturally to be pressureless.
This means that dark energy and dark matter energy densities must scale in time 
in a very different way. If their nearly equal magnitudes in the present day  universe   is
non accidental, this problem, known as "cosmic coincidence problem"\cite{coinc}
  appears 
to be one of the strongest 
challenge to both cosmology and particle physics.

The wish  to combine a possibility to describe an accelerating expansion 
for the present day universe with a solution of the cosmic
coincidence problem in the framework of a single consistent model,
was a main motivation of a number of recent attempts  to modify the
 particle physics models underlying the quintessence scenarios. The basic 
idea of these modifications consists in  specific changes of the fundamental 
(or effective) Lagrangian of the field theory intended to
 provide possibilities for "over-pumping"
of a part of  the  dark energy into the dark matter energy.
Cosmological consequences of a scalar field with exponential
potential coupled to matter was discussed in Refs.\cite{Wett},\cite{G-87}.
One of the popular  modifications of the particle physics model is known 
as variable-mass particles (VAMPs), an idea which was discussed in
papers of Refs.\cite{G-87},\cite{VAMP1}. One of the aims of authors of 
Ref.\cite{VAMP1} was to solve the age problem
of the Universe. In subsequent papers it was then realized that
 the desirable  effect for the coincidence problem
simultaneously with an accelerated expansion can be
achieved by assuming that mass of the dark matter particle depends on the
(quintessence) scalar field in such a way that
the cosmic time evolution of the quintessence field
is accompanied by increasing the mass of the dark matter particles. By 
 tuning of the appropriate parameters one can achieve both accelerated 
expansion and cosmic coincidence. This and other possible realizations
 of the basic idea during the last three years of the intensive study
of this intriguing problem
are listed partially in Refs.\cite{VAMP2}-\cite{VAMP9}.

Together with a definitely successful fitting of some of the observational data, 
the most
significant   achievement of these models consists in the possibility of 
providing both the accelerating expansion and a  resolution of
the coincidence problem without fine tuning\footnote{See however the recent 
paper\cite{Franca}}.
In spite of this they still have 
fundamental problem.
 Although there are  some  justifications for choices of 
certain types of  
dark matter-dark energy coupling  in the Lagrangian,\footnote{Such 
justifications are based for example on Brans-Dicke type 
theories\cite{VAMP5}
 or on strongly coupled string theory\cite{Gasp-Ven}
or on the large extra dimensions 
models\cite{extra-dim} or
on the supersymmetric theories\cite{VAMP7}}
there is a necessity to 
assume\cite{break-universality},\cite{VAMP2},\cite{VAMP5},\cite{VAMP6}
the absence or extremely strong suppression of the barion matter-dark energy 
coupling\footnote{In terminology of Ref.\cite{VAMP5} it is called
"choice of dark-dark coupling" }.  Actually this problem was known from the
very beginning  in the 
quintessence models since generically there are no reasons for the absence of  a 
direct coupling of the quintessence scalar field $\phi$ to the barion matter.
Such coupling  would be the origin of a long range scalar force because
of the very small mass of the quintessence field  $\phi$. 
This ''fifth-force" problem
might be solved if there would be a shift symmetry $\phi\rightarrow\phi +const$ 
of the action\cite{Carroll} which should be a reason for a strong suppression of
the direct quintessence - barion matter coupling.
 However the quintessence potential itself does not 
possess this symmetry. The situation with the ''fifth-force" problem becomes 
still more critical in the discussed above models since one should explain now
why the direct quintessence-dark matter coupling is permissible in the Lagrangian
while the same
is forbidden for the barion matter.
 Irrespective of the nature of the dark matter,
it is very hard to believe that there may be
a symmetry which is responsible for such situation.

The above mentioned modifications of the particle physics models are based on
 the assumption
that   all the fields of  the fundamental particle theory should be divided into two 
large groups:  one describing  detectable particles (ordinary matter) and the other
including dark matter particles.  
The main purpose of this paper is to demonstrate that there is a field theory which 
is able to propose a resolution for the above problems by an absolutely new way:
{\bf the dark matter is not introduced as a special type of matter but rather it 
appears as the solution of equations of motion describing a new type of states}
 of the (primordial) neutrino field; in other
words, {\bf  the dark neutrino matter and the regular neutrino generations
 (electron, muon
and $\tau$ neutrinos) are different states of the same primordial fermion
field}.

This field theory is the Two Measures Theory (TMT) originally built with the aim to solve 
the "old" cosmological constant problem\cite{GK1}-\cite{G1} and
was not addressed to the above problems from the beginning. The TMT model we continue to study in the
 present paper possesses spontaneously broken global scale symmetry\cite{G1}
which includes the shift symmetry $\phi\rightarrow\phi +const$ and it
allows to suggest\cite{GK4},\cite{GK5} a simultaneous resolution both of  the
fermion families problem and of the ''fifth-force" problem. {\bf The theory
starts from one primordial fermion field for each type of leptons and 
quarks}, e.g., in $SU(2)\times U(1)$ gauge theory the fermion content is the
 primordial neutrino and electron fields and primordial $u$ and $d$ quark 
fields. It turns out that masses and interactions of fermions
as well as the structure of their contributions to the energy-momentum 
tensor depend on the fermion densities. 

At a  fermion energy density corresponding 
to  normal laboratory particle physics conditions (that we will call
"high fermion density"), the fermion energy-momentum tensor is canonical
and each of the primordial fermions splits into three different states
with different masses (one of these states should be realized via fermion
condensate). These states were identified in  Refs.\cite{GK4,GK5} with 
the mass eigenstates of the fermion 
generations, and this effect is treated as the families birth effect. 
Although the original action includes scale invariant interaction
of the dilaton field $\phi$ (playing the role of the quintessence field)
 with all primordial fermions, the effective  interaction of the dilaton
with the regular fermions of the first two generations appears
to be extremely suppressed.
 In other words, the interaction of the dilaton with matter observable
 in  gravitational experiments is
 practically switched off, and that solves the 
''fifth-force" problem.

In TMT, physics of fermions at very low densities turns out to be very different
from what we know in normal particle physics. The term very low fermion
density means here that the fermion 
energy density is comparable with the dark energy density. In this
case, in addition to the canonical contribution to the  energy-momentum 
tensor, each of the
primordial fermion fields has a noncanonical contribution in the form of a 
{\bf dynamical fermionic $\Lambda$   term}. 

If the fermion 
energy density is comparable with the dark energy density then as will be 
shown in the present paper, the theory predicts that the primordial fermion
may not split into  generations and in the FRW universe it can 
 participate in the expansion of the
universe by means of changing its own parameters. We call this effect
"Cosmo-Particle Phenomenon"  and refer to such states
as  Cosmo-Low Energy Physics (CLEP) states. 

As the first step in studying Cosmo-Particle Phenomena, in this paper 
we restrict ourselves to the consideration of a simplified cosmological 
model where universe  is filled by 
a homogeneous scalar field $\phi$ and uniformly distributed 
non-relativistic (primordial) neutrinos and antineutrinos in CLEP states.    
Such CLEP-neutrino matter is  detectable practically only
through gravitational interaction and this is why it can be regarded 
as a model of dark matter. The mass of CLEP-neutrino increases as $a^{3/2}$
where $a=a(t)$ is the scale factor.  This dark matter is also cold one in 
the sense that kinetic energy of neutrinos is negligible as compared to their
mass. However due to the dynamical fermionic $\Lambda$   term 
generated by neutrinos in CLEP state, this cold dark matter has negative
 pressure and its equation
of state approaches $p_{d.m.}=-\rho_{d.m.}$ as  $a(t)\rightarrow\infty$.
Besides, the energy density of this dark matter scales in a way very
similar to the dark energy which
includes both a cosmological constant and an exponential potential.
So, due to the Cosmo-Particle Phenomena, TMT allows the
universe to achieve both the 
accelerated expansion and cosmic coincidence without the need to
postulate the existence of a special sort of matter called "dark matter".
The remarkable feature of such
a Cosmo-Particle solution is that {\bf  the total energy density of the
universe
in this case is less than it would be in the universe free
of fermionic matter at all}. 

All these unusual effects of TMT are due to the existence of a
fundamental constraint that emerges as a consistency condition of the
equations of motion. According to the role of this constraint
it can be regarded as a new law of nature.

Some notions about the content and organization of the paper are needed.
Plenty of new ideas have obliged us to simplify the presentation of the 
theory as far as possible. For this reason we ignored gauge non-abelian
structure as well as Higgs fields and Higgs phenomena, although they were 
already presented in our earlier publication\cite{GK5}. We also ignore
here different possible ways for introducing neutrino mass term and describe
it in the simplest "naive" way because this item has no direct relation
to questions studied in this paper.
We are not concern also a number of important astrophysical features
of the Cosmo-Particle Phenomena, such as clustering properties of neutrinos in
CLEP states. As for the presentation of TMT itself (that requires 
explanation quite a number of ideas and calculations developed
in previous publications\cite{GK1}-\cite{GK3}), we have tried to
make the paper self-sufficient. Main ideas of TMT and some technical 
details one can 
find in Sec.II and in Appendixes B and C. Besides, in Sec.II and
in Appendix A
we present some brane models arguments in favor of the general
structure of the action in TMT.

\section{Main ideas of the Two Measures Theory}

The Two Measures Theory (TMT) is a generally coordinate invariant theory
where the action has the form
\begin{equation}
    S = \int L_{1}\Phi d^{4}x +\int L_{2}\sqrt{-g}d^{4}x
\label{S}
\end{equation}
 including two Lagrangians $ L_{1}$ and $L_{2}$ and two 
measures of the volume elements ($\Phi d^{4}x$ and $\sqrt{-g}d^{4}x$
respectively). One is the usual measure of integration
$\sqrt{-g}$ in the 4-dimensional space-time manifold equipped by the metric
 $g_{\mu\nu}$. Another is also a scalar density built of four scalar
fields
$\varphi_{a}$ ($a=1,2,3,4$)
\begin{equation}
\Phi
=\varepsilon^{\mu\nu\alpha\beta}\varepsilon_{abcd}\partial_{\mu}\varphi_{a}
\partial_{\nu}\varphi_{b}\partial_{\alpha}\varphi_{c}
\partial_{\beta}\varphi_{d}.
\label{Phi}
\end{equation}
Since $\Phi$ is a total derivative, a shift of $L_{1}$ by a
constant, $L_{1}\rightarrow L_{1}+const$, has no effect on the equations
of
motion. Similar shift of $L_{2}$ would lead to the change of the constant 
part of the Lagrangian coupled to the volume element $\sqrt{-g}d^{4}x $.
In the standard General Relativity (GR), this constant term is the 
cosmological constant. 
However in TMT the relation between the
constant
 term of $L_{2}$ and the physical cosmological constant is very non
trivial
(see \cite{GK3},\cite{G1},\cite{K},\cite{G-O}).   

It is assumed that the Lagrangians $ L_{1}$ and $L_{2}$  are functions
of the matter
fields, the dilaton field, the metric, the connection  (or spin-connection )
 but not of the
"measure fields" $\varphi_{a}$. In such a case, i.e. when the 
measure fields $\varphi_{a}$ enter in the theory only via the measure
$\Phi$,
  the action (\ref{S}) has
the infinite dimensional symmetry\cite{GK3}:
$\varphi_{a}\rightarrow\varphi_{a}+f_{a}(L_{1})$, where $f_{a}(L_{1})$
is an arbitrary function of  $L_{1}$. 

Varying the measure fields $\varphi_{a}$, we get
\begin{equation}
B^{\mu}_{a}\partial_{\mu}L_{1}=0   
\label{dl0}
\end{equation}
where 
\begin{equation}
B^{\mu}_{a}=\varepsilon^{\mu\nu\alpha\beta}\varepsilon_{abcd}
\partial_{\nu}\varphi_{b}\partial_{\alpha}\varphi_{c}
\partial_{\beta}\varphi_{d}.
\label{Ama}
\end{equation}
Since
$Det (B^{\mu}_{a})
= \frac{4^{-4}}{4!}\Phi^{3}$ it follows that if $\Phi\neq 0$,
\begin{equation}
 L_{1}=sM^{4} =const
\label{varphi}
\end{equation}
where $s=\pm 1$ and $M$ is a constant of integration with the dimension of
mass.

Notice the {\bf  very important difference of TMT from scalar-tensor
theories with nonminimal coupling}: if the field $\Phi$ were the fundamental
(non composite) one then the variation of
$\Phi$ would results in the equation $L_{1}=0$.

Important feature of TMT that is responsible for many
interesting and desirable results of the field theory models studied 
so far\cite{GK1}-\cite{G-O}
 consists of the assumption that all fields, including 
also metric, connection (or vierbein and spin-connection) and the
{\bf measure fields} $\varphi_{a}$ are independent dynamical variables. All the 
relations between them are results of equations of motion. Such 
understanding
of the variational principle may be regarded as a generalization of the 
usual variational principle to the case when theory contains two measures
of integration and two Lagrangians. In particular, the independence of 
the metric and the connection means that we proceed in the first order 
formalism
and the relation between connection and metric is not necessarily 
according to Riemannian geometry. 
Applying the Palatini formalism in TMT one can show (see for example
\cite{GK3} or Appendix C of the present paper)  that 
the resulting relation between metric and  connection includes also the
gradient of the ratio of the two measures 
\begin{equation}
\zeta \equiv\frac{\Phi}{\sqrt{-g}}
\label{zeta}
\end{equation}
which is a scalar field. By an appropriate change of the dynamical 
variables which includes a conformal transformation of the metric,
one can formulate the theory as that in a Riemannian 
(or Riemann-Cartan) space-time. The corresponding conformal frame
we call "the Einstein frame". The big advantage of TMT is that in the 
very wide class of models, {\em the equations of motion
 take the canonical 
form of those of GR}, including the field theory models in curved 
space-time. One has to stress that {\bf TMT 
 has nothing common with any sort of the familiar scalar-tensor
theories}. All novelty of TMT in the
Einstein frame as compared with the standard GR is revealed only
 in an unusual structure of the 
effective potentials and interactions.  The key role in this effect 
belongs to the  scalar field  $\zeta (x)$.
Namely, the consistency condition of equations  
of motion has the form of algebraic constraint which determines $\zeta (x)$
as a function of matter fields. The surprising feature of the theory is
 that neither Newton constant nor curvature appears in this constraint
which means that the {\it geometrical scalar field} $\zeta (x)$
is determined by the matter fields configuration  locally and
straightforward (that is without gravitational interaction). As a result
of this, $\zeta (x)$ has a decisive influence in the determination of
the effective (that is appearing in the Einstein frame) interactions and
particle masses, and due to this, in the particle physics, cosmology 
and astrophysics.

As for the possible origin of the modified measure one can 
think of the following arguments.
Let us start by noticing that the modified measure part of the action
(\ref{S}), $\int L_{1}\Phi d^{4}x$, in the case $ L_{1}=const$
becomes a topological contribution to the action, since $\Phi$
is a total derivative.
It is very interesting that this structure can be obtained by 
considering the topological theory which results from 
studying "space-time filling branes", Refs.\cite{Zai}-\cite{Aka}.

Following for example Ref.\cite{Flor}, the Nambu-Goto action of a
3-brane embedded in a $4$-dimensional space-time\footnote{It should 
be pointed out that string and brane theories can be formulated 
with the use of a modified measure\cite{Mstring} giving some
new interesting results. In the spirit of the interpretation
developed in this section, such strings or branes could be 
regarded as extended objects moving in an embedding extended
object of the same dimensionality.} 
\begin{equation}
S_{NG}=T\int d^{4}x\sqrt{|\det (\partial_{\mu}\varphi^{a}
\partial_{\nu}\varphi^{b}G_{ab}(\varphi) |}
\label{NG}
\end{equation}
(where $G_{ab}(\varphi)$ is the metric of the embedding space),
is reduced to
\begin{equation}
\frac{T}{4!}\int d^{4}x
\varepsilon^{\mu\nu\alpha\beta}\varepsilon_{abcd}\partial_{\mu}\varphi_{a}
\partial_{\nu}\varphi_{b}\partial_{\alpha}\varphi_{c}
\partial_{\beta}\varphi_{d}
\sqrt{|\det G_{ab}(\varphi^{a})|}=
\frac{T}{4!}\int \Phi \sqrt{|\det G_{ab}(\varphi^{a})|}d^{4}x.
\label{Sreduced}
\end{equation}
 where $\Phi$ is determined by Eq.(\ref{Phi}).
The Euler-Lagrange equations for $\varphi^{a}$ are identities
which is the result of the fact\cite{Flor} that the action
(\ref{Sreduced}) is topological.  

This is the action of a "pure" space-filling brane governed just
by the identically constant brane tension. Let us now assume that 
the 3-brane is equipped with its own metric $g_{\mu\nu}$ and 
connection.
If we want to describe a 
 space-filling brane where gravity and matter are included
by means of the Lagrangian $ L_{1}=
L_{1}(g_{\mu\nu},connection,matter fields)$, we are led to the 
following form of the brane contribution 
to the action 
\begin{equation}
S_{1}= 
\int L_{1}\Phi \sqrt{|\det G_{ab}(\varphi^{a})|}d^{4}x
\label{S-red-NG-L1}
\end{equation}
Recall that $\sqrt{|\det G_{ab}(\varphi^{a})|}$ is a brane scalar. 
 
If $ L_{1}$ is not
identically a constant (we have seen above that 
it may become
a constant on the "mass shell", i.e. when equations of motion are 
satisfied),
then we are not talking any more of a topological contribution to 
the action.
Assuming again that $ L_{1}$ is 
$\varphi^{a}$ independent, it is easy to 
check that in spite of emergence of an additional factor 
$\sqrt{|\det G_{ab}(\varphi^{a})|}$ in Eq.(\ref{S-red-NG-L1})
as compared with the first term in Eq.(\ref{S}),
variation of $\varphi^{a}$ yields exactly the same 
equation (\ref{varphi}) if $\Phi\neq 0$. 
Moreover, all other equations of motion of the two measures
theory remains unchanged. The only effect of this 
additional factor consists in the redefinition of the
scalar field $\zeta$, Eq.(\ref{zeta}), where 
$\sqrt{|\det G_{ab}(\varphi^{a})|}$ emerges
as an additional factor.
Notice that the same results
are obtained if instead of $\sqrt{|\det G_{ab}(\varphi^{a})|}$ in 
Eq.(\ref{S-red-NG-L1}) there will be arbitrary function of $\varphi^{a}$. 

More arguments in favor of the general structure of the action
in TMT are given in Appendix A.

\section{Simplified scale-invariant model}

The TMT models possessing a global scale invariance\cite{G1,GK4,GK5}
are of significant interest because they demonstrate the possibility of  spontaneous 
breakdown of the scale symmetry\footnote{The field theory models with
explicitly broken scale symmetry and their application to the 
quintessential inflation type  cosmological scenarios have been studied
in Ref.\cite{K}. Inflation and transition to slowly accelerated phase 
from higher curvature terms was studied in Ref.\cite{G-O}. }. This effect 
appears due to the fact that the measure $\Phi$ is a total derivative.
 One of the 
interesting applications of the scale invariant TMT models\cite{GK4} is a possibility
to generate the exponential potential for the scalar field $\phi$ 
by means of the mentioned spontaneous symmetry breaking without introducing any potentials of  $\phi$ 
in the Lagrangians  $ L_{1}$ and $L_{2}$ in the action (\ref{S}). Some 
cosmological applications of this effect have been also studied in
 Ref.\cite{GK4}. Another application\cite{GK4,GK5}, rather surprising one,
 is the
possibility to suggest the way for the resolution of the fermion families 
problem and at the same time to solve the fifth force
problem.

The matter content of the simplified model we study here includes the scalar 
field $\phi$, two 
so-called primordial fermion fields (the neutrino primordial field $\nu$
 and the electron primordial field $E$) and electromagnetic field $A_{\mu}$.
The latter is included in order to demonstrate that gauge fields 
have no direct relation to the effects studied in this paper. 
Generalization to the non-Abelian 
gauge models including Higgs fields and quarks is straightforward\cite{GK5}.
To simplify the presentation of the features of TMT studied in this 
paper we ignore also the chiral properties of neutrino; this
can be done straightforward and does not affect the results 
obtained in this paper.

Keeping the general structure 
(\ref{S}), it is convenient to represent the action in the following 
form:
\begin{eqnarray}
S &=& \int d^{4}x e^{\alpha\phi /M_{p}}
(\Phi +b\sqrt{-g})\left[-\frac{1}{\kappa}R(\omega ,e)
+\frac{1}{2}g^{\mu\nu}\phi_{,\mu}\phi_{,\nu}\right]
-\int d^{4}x e^{2\alpha\phi /M_{p}}[\Phi V_{1} +\sqrt{-g}
V_{2}]
\nonumber\\
&+&\int d^{4}x e^{\alpha\phi /M_{p}}(\Phi +k\sqrt{-g})\frac{i}{2}
\left[\overline{\nu}
\left(\gamma^{a}e_{a}^{\mu}\overrightarrow{\nabla}_{\mu}
-\overleftarrow{\nabla}_{\mu}\gamma^{a}e_{a}^{\mu}\right)\nu
+\overline{E}
\left(\gamma^{a}e_{a}^{\mu}\overrightarrow{\nabla}^{(E)}_{\mu}
-\overleftarrow{\nabla}^{(E)}_{\mu}\gamma^{a}e_{a}^{\mu}\right)E
\right]
\nonumber\\
&-&\int d^{4}x e^{\frac{3}{2}\alpha\phi /M_{p}}
\left[(\Phi +h_{\nu}\sqrt{-g})\mu_{\nu}\overline{\nu}\nu
+(\Phi +h_{E}\sqrt{-g})\mu_{E}\overline{E}E
\right]
-\int d^{4}x\sqrt{-g}
\frac{1}{4}g^{\alpha\beta}g^{\mu\nu}F_{\alpha\mu}F_{\beta\nu}
\label{totaction}
\end{eqnarray}
where  $V_{1}$ and $V_{2}$ are constants, 
$F_{\alpha\beta}=\partial_{\alpha}A_{\beta}-
\partial_{\beta}A_{\alpha}$, \quad
   $\mu_{\nu}$ and
$\mu_{E}$ are  the mass parameters,
\\
$\overrightarrow{\nabla}_{\mu}=\overrightarrow{\partial}_{\mu}+
\frac{1}{2}\omega_{\mu}^{cd}\sigma_{cd}$,
$\overleftarrow{\nabla}_{\mu}=\overleftarrow{\partial}_{\mu}-
\frac{1}{2}\omega_{\mu}^{cd}\sigma_{cd}$,
$\overrightarrow{\nabla}^{(E)}_{\mu}=\overrightarrow{\partial}_{\mu}+
\frac{1}{2}\omega_{\mu}^{cd}\sigma_{cd}+ieA_{\mu}$ and
\\
$\overleftarrow{\nabla}^{(E)}_{\mu}=\overleftarrow{\partial}_{\mu}-
\frac{1}{2}\omega_{\mu}^{cd}\sigma_{cd}-ieA_{\mu}$;
$R(\omega ,e) =e^{a\mu}e^{b\nu}R_{\mu\nu ab}(\omega)$ is
the scalar curvature,
\\
 $e_{a}^{\mu}$ and
$\omega_{\mu}^{ab}$ are the vierbein  and spin-connection
respectively;
$g^{\mu\nu}=e^{\mu}_{a}e^{\nu}_{b}\eta^{ab}$ and 
\begin{equation}
R_{\mu\nu ab}(\omega)=\partial_{\mu}\omega_{\nu ab}   
+\omega_{\mu a}^{c}\omega_{\nu cb}
-(\mu\leftrightarrow\nu).
        \label{B}
\end{equation}

The action (\ref{totaction}) is invariant
under
the global scale transformations
\begin{eqnarray}
    e_{\mu}^{a}\rightarrow e^{\theta /2}e_{\mu}^{a}, \quad
\omega^{\mu}_{ab}\rightarrow \omega^{\mu}_{ab}, \quad
\varphi_{a}\rightarrow \lambda_{a}\varphi_{a}\quad
where \quad \Pi\lambda_{a}=e^{2\theta}
\nonumber
\\
A_{\alpha}\rightarrow A_{\alpha}, \quad
\phi\rightarrow \phi-\frac{M_{p}}{\alpha}\theta ,\quad
\Psi_{i}\rightarrow e^{-\theta /4}\Psi_{i}, \quad
\overline{\Psi}_{i}\rightarrow  e^{-\theta /4} \overline{\Psi}_{i}
\label{stferm}
\end{eqnarray}
where $\Psi_{i}$ ($i=\nu , E$) is the
general notation for the primordial fermion fields $\nu$ and $E$.

In (\ref{totaction}) there are two types of the gravitational terms and
 of the "kinetic-like terms" (both 
for the scalar field and for the primordial fermionic ones) which
respect the scale invariance : the terms of the one type coupled to the  
 measure $\Phi$ and those of the other type 
coupled to the measure 
$\sqrt{-g}$.
One should point out also the possibility of introducing two different 
 potential-like exponential functions coupled to the measures   
$\Phi$ and $\sqrt{-g}$ with constants $V_{1}$ and $V_{2}$ 
respectively\footnote{In more realistic models instead of these constants there 
will be functions of the Higgs field 
(see Ref.\cite{GK5})} . For the same reason there are two 
different sets of the mass-like terms of the primordial fermions. 
Constants $b, k, h_{\nu}, h_{E}$ are non specified dimensionless real parameters
of the model and we will only assume that {\em all these parameters are of 
the same order of magnitude}. The real positive parameter $\alpha$ is
assumed to be of the order of one.

The choice of the action (\ref{totaction})
needs a few additional explanations:

1) In order to avoid a possibility of negative energy contribution
from the space-time derivatives of the dilaton $\phi$ (see Ref.\cite{GK4}) we 
have chosen the
coefficient $b$ in front of $\sqrt{-g}$ in the first
integral of (\ref{totaction}) to
be a common factor of the gravitational term
 $-\frac{1}{\kappa}R(\omega ,e)$ and of the kinetic term for $\phi$. This
 guarantees that 
this item can not be an
origin of ghosts in quantum theory. 

2) For the same reasons we choose
the kinetic term of $A_{\mu}$ in the conformal invariant   
form which is possible if this term is coupled  only to the
measure $\sqrt{-g}$. Introducing the coupling of this term to
the measure $\Phi$
would lead to a nonlinear equation and non positivity of the
energy.

3) One can show that  in more realistic theories, like electro-weak one,
the right chiral structure in the Einstein frame demands the coupling of
 the kinetic terms of all the left and right primordial fermions 
to the measures
to be universal. In our simplified version of the theory, this feature 
is displayed in the choice of the parameter $k$  to be the common factor in 
front of the corresponding kinetic terms of the both primordial fermions.

Except for these three items, Eq.(\ref{totaction}) describes {\bf
the most
general TMT action satisfying the formulated above symmetries}.

\section{Constraint and equations of motion in the
Einstein frame}

Variation of
the measure fields $\varphi_{a}$ with the condition
$\Phi\neq 0$ leads, as we have already seen
in Sec.II, to the equation $ L_{1}=sM^{4}$  
where $L_{1}$ is now defined, according to  Eq. (\ref{S}), as
the part of the integrand of the action (\ref{totaction})
coupled to the measure $\Phi$. 
It can be noticed that the appearance of a nonzero integration
constant $sM^{4}$ spontaneously breaks the scale invariance 
(\ref{stferm}). The explicit form of the equation $ L_{1}=sM^{4}$
is presented in Appendix B, Eq.(\ref{app1}) as well as all other
equations of motion. In Appendix C one can find explicit solution
for the spin-connection.
One can see that the measure $\Phi$ degrees of freedom 
appear in all the equations of motion only through dependence
on the scalar 
field $\zeta $.
In particular, the gravitational and all matter
fields equations of motion include noncanonical terms 
proportional to $\partial_{\mu}\zeta$.

It turns out that with the set of the new variables ($\phi$ and 
$A_{\mu}$ remain the same)
\begin{equation}
\tilde{g}_{\mu\nu}=e^{\alpha\phi/M_{p}}(\zeta +b)g_{\mu\nu}, \quad  
\tilde{e}_{a\mu}=e^{\frac{1}{2}\alpha\phi/M_{p}}(\zeta
+b)^{1/2}e_{a\mu}, \quad
\nu^{\prime}=e^{-\frac{1}{4}\alpha\phi/M_{p}}
\frac{(\zeta +k)^{1/2}}{(\zeta +b)^{3/4}}\nu , \quad
E^{\prime}=e^{-\frac{1}{4}\alpha\phi/M_{p}}
\frac{(\zeta +k)^{1/2}}{(\zeta +b)^{3/4}}E
\label{ctferm}
\end{equation}
which we call the Einstein frame,
 the spin-connections (\ref{C4}) become those of the 
Einstein-Cartan space-time and the noncanonical terms proportional to 
$\partial_{\mu}\zeta$ disappear from all equations of motion. Since  
$\tilde{e}_{a\mu}$,
$\nu^{\prime}$ and $E^{\prime}$ are invariant under the scale 
transformations
(\ref{stferm}), spontaneous breaking of the scale symmetry (\ref{stferm})
(by means of Eq.(5)) is reduced in the new variables to the
{\bf spontaneous breaking of the shift symmetry} 
$\phi\rightarrow\phi +const$. 

The gravitational 
equations (\ref{app12}) after the change of variables (\ref{ctferm}) to 
the Einstein frame and some simple algebra, take the standard GR form
\begin{equation}
G_{\mu\nu}(\tilde{g}_{\alpha\beta})=\frac{\kappa}{2}T_{\mu\nu}^{eff}
 \label{gef}
\end{equation}
\begin{equation}
T_{\mu\nu}^{eff}=\phi_{,\mu}\phi_{,\nu}-\frac{1}{2}
\tilde{g}_{\mu\nu}\tilde{g}^{\alpha\beta}\phi_{,\alpha}\phi_{,\beta}
+\tilde{g}_{\mu\nu}V_{eff}(\phi ;\zeta)+T_{\mu\nu}^{(em)}
+T_{\mu\nu}^{(ferm,can)}+T_{\mu\nu}^{(ferm,noncan)}
 \label{Tmn}
\end{equation}
Here $G_{\mu\nu}(\tilde{g}_{\alpha\beta})$ is the Einstein tensor
in the Riemannian space-time with the metric
$\tilde{g}_{\mu\nu}$; the function $V_{eff}(\phi ;\zeta)$ has the form 
\begin{equation}
V_{eff}(\phi ;\zeta)=
\frac{b\left(sM^{4}e^{-2\alpha\phi/M_{p}}+V_{1}\right)-V_{2}}{(\zeta
+b)^{2}};
\label{Veff1}
\end{equation}
$T_{\mu\nu}^{(em)}$ is the canonical energy momentum tensor for the
electromagnetic field;
$T_{\mu\nu}^{(ferm,can)}$ is the canonical energy momentum tensor for
(primordial) fermions $\nu^{\prime}$ and $E^{\prime}$ in
curved space-time
including also standard electromagnetic interaction of $E^{\prime}$.
$T_{\mu\nu}^{(ferm,noncan)}$ is the {\em noncanonical} contribution
of the fermions into the energy momentum tensor 
\begin{equation}
 T_{\mu\nu}^{(ferm,noncan)}=-\tilde{g}_{\mu\nu}[F_{\nu}(\zeta)
\overline{\nu^{\prime}}\nu^{\prime}+
F_{E}(\zeta)
\overline{E^{\prime}}E^{\prime}]\quad\equiv\quad
\tilde{g}_{\mu\nu}\Lambda_{dyn}^{(ferm)}
\label{Tmn-noncan}
\end{equation}
where
\begin{equation}
F_{i}(\zeta)\equiv
\frac{\mu_{i}}{2(\zeta +k)^{2}(\zeta +b)^{1/2}}
[\zeta^{2}+(3h_{i}-k)\zeta +2b(h_{i}-k)+kh_{i}], \qquad 
i=\nu^{\prime},E^{\prime}.
 \label{Fizeta1}
\end{equation}
The structure of 
$T_{\mu\nu}^{(ferm,noncan)}$ shows that it behaves as
 a sort of  variable cosmological constant\cite{var-Lambda}
but in our case it is originated by fermions. This is
why we will refer to it as {\bf dynamical fermionic $\Lambda$  
 term}. This fact is displayed explicitly 
in Eq.(\ref{Tmn-noncan}) by defining
$\Lambda_{dyn}^{(ferm)}$. One has to 
emphasize
the substantial difference of the way $\Lambda_{dyn}^{(ferm)}$
 emerges here as compared to the models of the
condensate cosmology (see for example Refs.\cite{cond1}-\cite{cond3}).
In those models the dynamical cosmological constant resulted from
 bosonic or fermionic condensation. In TMT model studied here, 
 $\Lambda_{dyn}^{(ferm)}$ is 
originated by fermions but there is no need in any condensate.  
As we will see in the next sections, 
$\Lambda_{dyn}^{ferm}$ becomes negligible in gravitational experiments
with observable matter.
However it may be  very important for some
astrophysics and cosmology problems and in particular it plays
a key role in the resolution of the cosmic coincidence
and other  problems discussed in the
 Introduction.

The dilaton $\phi$ field equation (\ref{app13}) in the new variables
reads
\begin{equation}
\Box\phi -\frac{\alpha}{M_{p}(\zeta +b)} 
\left[sM^{4}e^{-2\alpha\phi/M_{p}}-\frac{(\zeta -b)V_{1}+2V_{2}}{\zeta
+b}\right]=
-\frac{\alpha }{M_{p}}[F_{\nu}\overline{\nu^{\prime}}\nu^{\prime}+
F_{E}\overline{E^{\prime}}E^{\prime}],
\label{phief+ferm1}
\end{equation}
where $\Box\phi =(-\tilde{g})^{-1/2}\partial_{\mu}
(\sqrt{-\tilde{g}}\tilde{g}^{\mu\nu}\partial_{\nu}\phi)$.

Equations (\ref{app7}),(\ref{app8}) for the primordial leptons 
in terms of the
variables (\ref{ctferm}) take the standard form of fermionic equations
for $\nu^{\prime}$, $E^{\prime}$
in the Einstein-Cartan space-time  where the standard electromagnetic
interaction 
 presents also. All the novelty consists of the form of
the $\zeta$ depending "masses" $m_{i}(\zeta)$ of the primordial fermions
$\nu^{\prime}$, $E^{\prime}$:
\begin{equation}
m_{i}(\zeta)=
\frac{\mu_{i}(\zeta +h_{i})}{(\zeta +k)(\zeta +b)^{1/2}}
\qquad i=\nu^{\prime},E^{\prime}.
 \label{muferm1} 
\end{equation} 
It should be noticed that change of variables we have performed
by means of Eq.(\ref{ctferm}) provide also a conventional form
of the covariant conservation law of fermionic current
$j^{\mu}=\overline{\Psi}^{\prime}\gamma^{a}\tilde{e}_{a}^{\mu}\Psi^{\prime}$. 

The scalar field $\zeta$ in the above equations is defined
by the constraint determined by means of Eqs. (\ref{app10}),(\ref{app11}) which
in the new variables
(\ref{ctferm}) takes the form
\begin{equation}
-\frac{1}{(\zeta +b)^{2}}\left\{(\zeta
-b)\left[sM^{4}e^{-2\alpha\phi/M_{p}}+
V_{1}\right]+2V_{2}\right\}=
F_{\nu}(\zeta)\overline{\nu^{\prime}}\nu^{\prime}+
F_{E}(\zeta)\overline{E^{\prime}}E^{\prime}\quad \equiv \quad
-\Lambda_{dyn}^{(ferm)}
\label{constraint3}
\end{equation}
Notice that neither electromagnetic field nor kinetic terms enter 
into the constraint. One should point out the interesting and very 
important fact: the same functions $F_{i}(\zeta)$ 
($i=\nu^{\prime},E^{\prime}$) emerge in 
 $\Lambda_{dyn}^{(ferm)}$, in the effective coupling of the dilaton
to fermions and in the constraint.

Applying constraint (\ref{constraint3}) to
Eq.(\ref{phief+ferm1}) one can reduce the latter to the form     
\begin{equation}
\Box\phi-\frac{2\alpha\zeta}{(\zeta +b)^{2}M_{p}}sM^{4}e^{-2\alpha\phi/M_{p}}
=0,
\label{phi-constr}
\end{equation}
where $\zeta$  is a solution of the constraint (\ref{constraint3}). 

Generically the constraint (\ref{constraint3}) determines $\zeta$ as 
a very complicated 
function of $\phi$, \,
$\overline{\nu^{\prime}}\nu^{\prime}$ and $\overline{E^{\prime}}E^{\prime}$ . 
Substituting
the appropriate solution for $\zeta$ into the above equations
of motion one can conclude that
in general, there is no sense, for example, to regard 
 $V_{eff}(\phi ;\zeta)$, Eq.(\ref{Veff1}), as the effective 
potential for the scalar field $\phi$ because it depends in a very
nontrivial way on $\overline{\nu^{\prime}}\nu^{\prime}$ and
 $\overline{E^{\prime}}E^{\prime}$ as well.
For the same reason, the dynamical fermionic  
$\Lambda_{dyn}^{(ferm)}$ term, as well as
the effective fermion "mass", Eq.(\ref{muferm1}), 
describe, in general,  self-interactions of the primordial fermions
depending also on the 
scalar field $\phi$. Therefore {\em it is impossible, in
general, to separate the terms of} $T_{\mu\nu}$ {\em which describe
 the scalar field $\phi$ effective potential from the fermion 
contributions}. Such mixing of the scalar field $\phi$ and 
fermionic matter gives rise to a rather complicated system 
of equations when trying 
to apply the theory to a general situations that can appear 
in particle physics, 
astrophysics and
cosmology. 

One can note, however, that if for some reasons $\zeta$ occurs
to be constant of the order of the parameter $b$ 
then  the l.h.s. of the constraint (\ref{constraint3}) 
is of the same order of
magnitude as $V_{eff}(\phi ;\zeta)$, Eq.(\ref{Veff1}),
 which in this case is the effective potential of the dilaton field
$\phi$. We will see that due to our assumption that all the dimensionless
parameters
in the action (\ref{totaction}) are of the same order of magnitude, 
 $\zeta$ is in fact of the 
order of those parameters in the relevant cosmological solutions studied
below. 
 So, roughly speaking, {\bf  the constraint  
describes a balance}  ( in orders of magnitude) which must be realized,
 according
to TMT, {\bf between the dark energy density, on the one hand, and 
the dynamical fermionic $\Lambda_{dyn}^{(ferm)}$ term,
 on the other hand}.
One has to remind here that the constraint dictating this balance is 
a consistency condition of equations of motion, from the  
mathematical point of view.  From the physics point of view,
the constraint emerges as a new law of nature\footnote{It is interesting 
to note that on the basis of the observational data, 
general arguments have been given\cite{Chernin}  in favor of the existence of a
fundamental principle that correlates somehow different contributions
to the energy density of the universe. }. 

There are  a few very 
important particular situations where the theory allows exact
solutions of great interest\cite{GK4},\cite{GK5}. We will see that
for the late time universe, the constraint provides 
a possibility to explain the cosmic coincidence.

\section{Dark energy in the absence of massive fermions}

In the case of the complete absence of massive fermions the constraint determines
$\zeta$ as the function of $\phi$   
\begin{equation}
\zeta = b-\frac{2V_{2}}
{V_{1}+sM^{4}e^{-2\alpha\phi/M_{p}}}.
\label{zeta-without-ferm}
 \end{equation}
The effective potential of the scalar field $\phi$ results then from Eq.(\ref{Veff1})
\begin{equation} 
V_{eff}^{(0)}(\phi)\equiv V_{eff}(\phi;\zeta)|_{\overline{\psi^{\prime}}\psi^{\prime}=0}
=\frac{(V_{1}+sM^{4}e^{-2\alpha\phi/M_{p}})^{2}}
{4[b(V_{1}+sM^{4}e^{-2\alpha\phi/M_{p}})-V_{2}]}
\label{Veffvac}    
\end{equation}
and the $\phi$-equation (\ref{phief+ferm1}) is reduced to
\begin{equation}
\Box\phi +V^{(0)\prime}_{eff}(\phi)=0,
\label{eq-phief-without-ferm}
\end{equation}
where prime sets derivative with respect to $\phi$.

Applying this model to the cosmology of the late time universe
and assuming  that the scalar field $\phi\rightarrow\infty$ as
$t\rightarrow\infty$, we see that   
the evolution of the late time universe  is governed by the
 sum of the
cosmological constant
\begin{equation}
\Lambda^{(0)}   
=\frac{V_{1}^{2}}
{4(bV_{1}-V_{2})}
\label{lambda-without-ferm}
\end{equation}
and the quintessence-like scalar field with the potential
\begin{equation}   
V^{(0)}_{quint}(\phi)
=\frac{(bV_{1}-2V_{2})sV_{1}M^{4}e^{-2\alpha\phi/M_{p}}+
(bV_{1}-V_{2})M^{8}e^{-4\alpha\phi/M_{p}}}
{4(bV_{1}-V_{2})[b(V_{1}+sM^{4}e^{-2\alpha\phi/M_{p}})-V_{2}]}.
\label{V-quint-without-ferm}
\end{equation}

The cosmological constant $\Lambda^{(0)}$ is the asymptotic value
( as $t\rightarrow\infty$) of the dark energy
density
\begin{equation}
\rho^{(0)}_{d.e}=\frac{1}{2}\dot{\phi}^{2}+\Lambda^{(0)}+V^{(0)}_{quint}(\phi).
\label{rho-without-ferm}
\end{equation}
for the FRW universe in the model where massive
fermions absent. $\Lambda^{(0)}$
is positive provided
\begin{equation} 
bV_{1}>V_{2}
\label{bv1>v2}
\end{equation}
that will be assumed in what follows.

There are two ways to provide the observable order of magnitude of
the vacuum energy density by an appropriate choice of the parameters  
of the theory.
\\

(a)
The first one consists of the assumption that the parameter
$b$ is of the order of unity; then there is no need for $V_{1}$  
and $V_{2}$ to be small: it is enough that $V_{2}<0$ and
 $V_{1}\ll |V_{2}| $.  This possibility is a kind of a {\bf seesaw} mechanism
(see \cite{G1}). For instance, if $V_{1}\sim (10^{3}GeV)^{4}$ and
$V_{2}\sim (10^{15}GeV)^{4}$ then $\Lambda^{(0)}\sim (10^{-3}eV)^{4}$.
\\

(b)
The second possibility is  to choose the dimensionless parameter $b>0$
to be very large.  In this case the order of magnitude of $V_{1}$
and $V_{2}$ could be either as in the above case (a) or to be  not too 
much different (or even of the same order).

The mass of a test fermion, that is the primordial fermion field in
a state having negligible contributions into equations (\ref{Tmn}), 
(\ref{phief+ferm1}), (\ref{constraint3}),
 is $\phi$-dependent as it is determined 
by Eqs.(\ref{muferm1}) and (\ref{zeta-without-ferm}). The asymptotic
value of the test fermion mass as $\phi\rightarrow\infty$ is very small
\begin{equation}
m_{i}^{(test)}|_{\zeta =b-2V_{2}/V_{1}}=
\frac{\mu_{i}}{\sqrt{2}}\frac{[(b+h_{i})V_{1}-2V_{2}]V_{1}^{1/2}}
{[(b+k)V_{1}-2V_{2}](bV_{1}-V_{2})^{1/2}} \qquad 
i=\nu^{\prime},E^{\prime}
 \label{m-test}
 \end{equation}
because it is proportional to $(\Lambda^{(0)}/V_{1})^{1/2}$.

\section{Regular fermion (lepton) families birth effect}

We are going here to demonstrate the effect of splitting of each type of
 the primordial fermions into three generations in
the regime  where fermion densities are of the typical laboratory
particle physics scales. This effect has been discussed  in 
Ref.\cite{GK5} for lepton and quark primordial fields
in the framework of models with massless neutrino.
Namely, when the fermion (in particular, neutrino)
primordial field is introduced into the theory in such a way that
it remains massless then it does not appear in the constraint. As a result,
in such model there is no splitting of the primordial neutrino into
three generations. In the present model the primordial neutrino field
is regarded on the same manner as the primordial electron field $E$. 
 
Consider the constraint (\ref{constraint3}) in a typical particle physics
situation, say detection of a single fermion, i.e. a free one particle state
of the field $\Psi^{\prime}$ (which in our simplified model can be either 
$E^{\prime}$ or $\nu^{\prime}$).  This measurement implies a 
localization of the fermion which is expressed in developing a very big 
value of  $|\overline{\Psi^{\prime}}\Psi^{\prime}|$ (for the cases when
$\zeta =const$, in the local  particle rest frame, 
$\overline{\Psi^{\prime}}\Psi^{\prime}$ is, as a matter of fact, the 
probability density). There are two ways to provide a balance dictated
by the constraint:

either
\begin{equation}
F_{i}(\zeta)\approx 0, \qquad i=\nu^{\prime},E^{\prime}
\label{F-approx-0}
\end{equation}

or
\begin{equation}
\zeta +b\approx 0
\label{z-approx-0}.
\end{equation}

Eq.(\ref{F-approx-0}) determines  two sets of the
constant solutions for $\zeta$ 
\begin{equation}
\zeta_{1,2}^{(i)}=\frac{1}{2}\left[k-3h_{i}\pm\sqrt{(k-3h_{i})^{2}+
8b(k-h_{i})
-4kh_{i}}\,\right], \qquad i=\nu^{\prime},E^{\prime}
\label{zetapm1}   
\end{equation}
They  correspond
to two different states of the primordial leptons with {\bf 
different constant 
masses} determined by Eq.(\ref{muferm1}) where we have to substitute
$\zeta_{1,2}^{(i)}$ instead of $\zeta$.
These two states can be identified with the mass eigenstates of
{\em the first two generations of the 
regular leptons}.

Comparing the simplified scale invariant model we have restricted ourselves
 in the present paper with
the more realistic non Abelian gauge invariant  models\cite{GK5} including
Higgs fields, one can see that the parameters $V_{1}$ and $V_{2}$ in the 
action (\ref{totaction}) have a sense of the Higgs field potentials
contributing
to  the constraint and to  the dark energy density.
On the basis of the  above analysis we conclude
that each type of {\bf the primordial fermion fields} (in the present model:
$\nu^{\prime}$ and $E^{\prime})$ {\bf turn into regular fermions
of the first two generations, only if
its canonical energy density} $\tilde{g}^{00}T_{00}^{(ferm,can)}$ 
{\bf is much greater than} $\Lambda_{dyn}^{(ferm)}\propto F_{i}(\zeta)\approx 0$ 
{\bf being 
(as it is dictated by the constraint) of the order of
 the typical contributions into
the dark  energy density from the dilaton and Higgs fields}. For
example, if the primordial neutrino is in the state with $\zeta =
\zeta^{(\nu)}_{1}$ (or $\zeta =\zeta^{(\nu)}_{2}$) determined by
Eq.(\ref{zetapm1}), it is detected as the regular
electron  neutrino $\nu_{e}$ (or as the muon neutrino $\nu_{\mu}$)
respectively. The states of the primordial electron field $E$ with
$\zeta =\zeta^{(E)}_{1}$ and $\zeta =\zeta^{(E)}_{2}$ are detected as
the regular electron $e$ and muon $\mu$ respectively. Similar
identification is true also for quarks\cite{GK5}. 

The third solution, Eq.(\ref{z-approx-0}), we associate with
the third fermion generation. It becomes  singular
if one takes it as an exact equality (this can be seen from equations 
of motion). This means that one can not neglect the rest of the terms 
in the constraint (\ref{constraint3}). Then 
by solving $\zeta +b$ in terms of $\phi$ and 
$\overline{\Psi^{\prime}_{i}}\Psi^{\prime}_{i}$
($i=\nu^{\prime},E^{\prime}$) we have approximately 
$\zeta=\zeta_{3}\approx -b$ and
\begin{equation}
\frac{1}{\sqrt{\zeta_{3} +b}}\approx
\left[\frac{\mu_{i}(b-h_{i})\overline{\Psi^{\prime}_{i}}\Psi^{\prime}_{i}}
{4(b-k)\left[b\left(sM^{4}e^{-2\alpha\phi/M_{p}}+V_{1}\right)-V_{2}
\right]}\right]^{1/3}, \qquad i=\nu^{\prime},E^{\prime}.
\label{srtzeta1}
\end{equation}
Inserting this into Eq.(\ref{muferm1} ) we see that the fermion
self-interaction term in the effective Lagrangian
\begin{equation}
L_{selfint.}^{(3-rd\, family)}\approx
\left[\frac{\mu_{i}^{4}(b-h_{i})^{4}(\overline{\Psi^{\prime}_{i}}\Psi^{\prime}_{i})^{4}}
{4(b-k)^{4}\left[b\left(sM^{4}e^{-2\alpha\phi/M_{p}}+V_{1}\right)-V_{2}
\right]}\right]^{1/3},
 \qquad i=\nu^{\prime},E^{\prime}.
\label{L-selfint-3}
\end{equation}
 is produced in the regime (\ref{z-approx-0}).
This fermion self-interaction is non-polynomial  with dimensionless
coupling constant. A full 
treatment of the third generation requires the study of quantum corrections 
and fermion condensates\cite{NJL},\cite{Nielsen} which will give the third
 fermion family appropriate masses. Notice that in contrast with the first two
generations, the noncanonical part of the primordial fermion
energy density in the regime (\ref{z-approx-0}) is non negligible:
 \begin{equation}
\Lambda_{dyn}^{ferm}|_{3-rd\, family}=-\frac{1}{2}L_{selfint.}^{(3-rd\, family)}
\label{3-rd-noncan}
\end{equation}

An interesting question in this theory is: how family mixing appears? 
Or in other words, what is the origin of the Cabbibo angle and more
generically of the Kobayashi-Maskawa mass matrix in this theory?
We now discuss one possible origin of mixing in TMT.

It is clear that the mass eigenstates defined by Eq.(\ref{zetapm1}) 
and the one defining the third generation, Eq.(\ref{srtzeta1})
(which does not get a mass at the classical level yet) are associated 
with configurations where a "pure $\nu$" state or a "pure $E$" state
, etc. are considered. Such one-particle states are achieved as 
free asymptotic states (far from interaction space-time region).
In contrast to this, in the interaction regions, different types of 
primordial fermions are present and therefore the values of $\zeta$
in these regions are different from those determined 
by Eqs.(\ref{zetapm1}) and (\ref{srtzeta1})
defining  the mass eigenstates. 
So, it follows that the fermion states participating at the vertices
of interaction in general {\em do not} coincide with the fermion
mass eigenstates and from this one can expect the existence of particle
 mixing, i.e. a Cabbibo angle or more generically, a Kobayashi-Maskawa
matrix with nondiagonal elements. This is the question of great
interest and it should be studied in the future developments of TMT.

\section{Resolution of the fifth force problem}

One can show that both in the cosmological scenario without fermions 
(as in Sec.V) and for the universe filled by fermionic matter
(see the next section), the mass of the dilaton (quintessence) field
$\phi$ is extremely small. Then we must explain why 
 there is no a direct coupling of the dilaton to the observed matter
(or why it is extremely suppressed), 
otherwise a long-range scalar force should be observed in
gravitational experiments in contradiction 
with the equivalence principle.

 According to the dilaton
equation (\ref{phief+ferm1}), there exists Yukawa-like coupling of the
primordial fermions to the dilaton with coupling "constants" 
$\frac{\alpha }{M_{p}}F_{i}(\zeta)$; this coupling becomes usual
Yukawa coupling when $\zeta =const$ .
Recall that to simplify the presentation of the theory we have not
included the quark primordial fields. In the complete formulation of
the theory  the quark primordial fields enter on the same manner as
leptonic. Therefore quarks also possess the mentioned Yukawa-like
coupling to the dilaton where the functions $F_{q}(\zeta)$ of the
 same form as Eq.(\ref{Fizeta1}) enter into the same form 
coupling "constants"  we have obtained for the dilaton-lepton
coupling. 

Similar to what was explained in the previous section,
the quark families birth effect occurs at high 
 magnitude of
$|\overline{\Psi}^{\prime}_{q}\Psi^{\prime}_{q}|$ 
corresponding to the laboratory conditions of the normal particle
physics: the balance dictated by the constraint requires then
either
$F_{q}(\zeta)\approx 0$ (i.e. $\zeta$ is a constant determined by
equations of the form of Eq.(\ref{zetapm1}) and corresponding to 
the first two quark generations)
 or $\zeta +b\approx 0$ (the third generation).

Practically all  fermionic matter observable in gravitational experiments
 consists of the regular quarks and leptons of the first two generations.
 In the context of TMT this means that only those primordial fermions
 participate in gravitational experiments which are in the state with
$F(\zeta)\approx 0$. Thus {\bf the Yukawa coupling "constant" of
the dilaton-fermion interactions observable in gravitational experiments
appears to be extremely strongly suppressed}\footnote{Note that the same
conclusion can be obtained considering the dilaton field equation
written in the form (\ref{phi-constr}). Indeed, the values of $\zeta$
corresponding to the first two generations are constants  and in 
particular they are $\overline{\Psi}^{\prime}_{q}\Psi^{\prime}_{q}$
independent with extremely high accuracy. Therefore substitution of 
the values of $\zeta$ into Eq.(\ref{phi-constr}) in this case
 does not produce dilaton-fermion coupling.}.
 The measure of the
suppression can be characterized by using the constraint (\ref {constraint3}): the fermionic
source in the dilaton equation  (\ref{phief+ferm1}),
$F(\zeta)\overline{\Psi^{\prime}}\Psi^{\prime}$, is of the order of
the dark energy density divided by $M_{p}$. 
This is the mechanism
of the resolution of the fifth force problem in TMT. Notice that this
mechanism does not suppress the interaction of the dilaton with the
primordial fermion fields if the latter are not in the state 
corresponding to  the first two generations. 

Recall that the fermionic contribution to the energy-momentum tensor
(\ref{Tmn}) consists of  two terms: canonical and noncanonical.
The latter is the dynamical fermionic $\Lambda_{dyn}^{(ferm)}$
term (proportional to $F(\zeta)$) which 
is of the order of the dark energy density.  This is why {\em in 
 gravitational experiments
with observable matter the noncanonical part of the 
fermionic contribution to the energy-momentum tensor is negligible
as compared to the canonical one}.

If the fermions of the third family would be a significant portion
of the  matter observable in 
 gravitational experiments then anomalous gravitational effects
could be detected. Indeed, the canonical part of the energy-momentum
tensor for the third family
fermion in the rest frame is
\begin{equation}
T_{\mu\nu}^{(ferm,can)}|_{3-rd\, family}=
\delta_{\mu 0}\delta_{\nu 0}L_{selfint.}^{(3-rd\, family)}
\label{can-3}
\end{equation}
whereas the correspondent  noncanonical part reads
\begin{equation}
T_{\mu\nu}^{(ferm,noncan)}=-\frac{1}{2}\tilde{g}_{\mu\nu}
L_{selfint.}^{(3-rd\, family)};
\label{noncan-3}
\end{equation}
the latter is not negligible and could be the origin of gravitational anomalies.

It is interesting that although $F_{i}(\zeta)\neq 0$ for the third fermion family,
its interaction with the dilaton turns out to be suppressed at the 
late-time universe. In fact, 
inserting  $\zeta$ corresponding to the third fermion family,
Eq.(\ref{srtzeta1}), into Eq.(\ref{phi-constr})
we see that the coupling of the dilaton
to the third fermion family includes the factor
$e^{-2\alpha\phi/M_{p}}$ which is extremely small in the
quintessential scenario of the late time universe. But this 
coupling might be important at earlier stages of the universe.

\section{Cosmo-Low Energy Physics states and cosmology of the late-time
universe}
\subsection{The essence of the Cosmo-Particle Phenomena}
\subsubsection{Toy model}
In the three previous sections we studied two opposite limiting
cases: one is realized if there are no  fermions at all or they appear 
only as test particles; the second
one corresponds to the laboratory conditions of the normal particle
physics where due to  localizations, the magnitude of
$|\overline{\Psi}^{\prime}\Psi^{\prime}|$ is so large
that balance dictated by the constraint requires either
$F_{i}(\zeta)\approx 0$  or $\zeta +b\approx 0$ .

It turns out that besides these  standard particle physics situations,
TMT predicts possibility of so far unknown
 states which can be realized,  for example,
in astrophysics and cosmology. 
To illustrate some of the properties of such states let us start
from a simplest (but non physical) model describing the following 
self-consistent system: the  FRW universe filled by the homogeneous
scalar field $\phi$ and 
a homogeneous primordial fermion field $\Psi^{\prime}
=\Psi^{\prime}(t)$ 
(one can think for instance of a primordial neutrino ) as sources
of gravity. The fermion has zero momenta and therefore one can 
rewrite $\overline{\Psi}^{\prime}\Psi^{\prime}$ in the form of
density $\overline{\Psi}^{\prime}\Psi^{\prime}= u^{\dagger}u$
where $u$ is the large component of the Dirac spinor $\Psi^{\prime}$. 
The space components of the 4-current 
$\tilde{e}_{a}^{\mu}\overline{\Psi}^{\prime}\gamma^{a}\Psi^{\prime}$  
equal zero. It follows from the 4-current conservation  that
$\overline{\Psi}^{\prime}\Psi^{\prime}= u^{\dagger}u
=\frac{const}{a^{3}}$ where $a=a(t)$ is the scale factor.
 It is convenient to rewrite the constraint 
(\ref{constraint3}) (adjusted to the 
model under consideration) 
in the form
\begin{equation}
\left\{\left[(\zeta -b)sM^{4}e^{-2\alpha\phi/M_{p}}+
V_{1}\right]+2V_{2}\right\}
+\frac{(\zeta +b)^{3/2}}{(\zeta +k)^{2}}
[\zeta^{2}+(3h_{\nu}-k)\zeta +2b(h_{\nu}-k)+kh_{\nu}]
\frac{const}{a^{3}}
=0,
\label{constraint-frw}
\end{equation}
where the function $F_{\nu}(\zeta)$, Eq.(\ref{Fizeta1}), has been used.  
 
A possible solution of the constraint for the expanding universe
as the scale factor 
$a(t)\rightarrow\infty$ is identical to the one studied 
in Sec.V
where the fermion contribution is treated as negligible.      
There is however another solution where the decaying fermion 
contribution $u^{\dagger}u\sim \frac{const}{a^{3}}$ to the 
constraint is compensated by
the appropriate behavior of the scalar field $\zeta$. Namely if
expansion of the universe is accompanied by
approaching $\zeta\rightarrow -k$ then the contribution of the last
term can be comparable with that of other terms at all times.
This regime corresponds to a very unexpected state of the primordial
fermion. First,
this state does not belong to any generation of the regular fermions.
Second, the effective mass of the fermion in this state increases
like $(\zeta +k)^{-1}$ and the behavior of
the dynamical fermionic $\Lambda_{dyn}^{(ferm)}$ term is
$\Lambda_{dyn}^{(ferm)}\propto(\zeta +k)^{-2}u^{\dagger}u$. 
This means that at
the late time universe, the canonical fermion energy density 
$\rho_{(ferm,canon)}\propto(\zeta +k)^{-1}u^{\dagger}u $
 becomes much less than 
 $\Lambda_{dyn}^{(ferm)}$. Third, such cold fermion matter
possesses pressure and its equation of state  in
the late time universe approaches the form 
$p_{(ferm)}=-\rho_{(ferm)}$.

   Since $\Lambda_{dyn}^{(ferm)}$ is of the order of
the dark energy density, 
we  conclude that for very low fermion energy densities
(of the order of the dark energy density),
TMT predicts a possibility of new type of states which we will
refer as "Cosmo-Low Energy Physics"  (CLEP) states. The effect
of changing parameters (like a mass) of the fermion in CLEP
state due to the described above its direct drawing 
into the  cosmological 
expansion will be referred as {\bf Cosmo-Particle Phenomenon}.

\subsubsection{Uniformly distributed non-relativistic neutrinos}

A more realistic model may be constructed by studying a possible
effect of uniformly distributed neutrinos (and antineutrinos)
on cosmological scenarios in the context of TMT. The simplest 
solution implies that primordial
 neutrino $\nu^{\prime}$ is regarded as
a free test fermion in the fermion vacuum (with $\zeta$ given 
by Eq.(\ref{zeta-without-ferm})) studied in Sec.V. Then
the mass of the neutrino is very small and its asymptotic value
(when assuming a quintessence-like scenario with 
$\phi\rightarrow\infty$ as $t\rightarrow\infty$) is 
determined by Eq.(\ref{m-test}). 

There is however a more
interesting solution motivated by what was discussed above in the
toy model.
 The important thing we have learned is
a possibility of  states where the primordial neutrinos
and antineutrinos can 
be very heavy if $\zeta\approx -k$. This is why it may have 
sense to consider  wave packets of 
free {\em non-relativistic neutrinos and antineutrinos},
 i.e. with the momentum $|\vec{p}_{\nu}|\ll m_{\nu}$.
We will assume also that the momentum-space widths of
the wave packets are very narrow ($\Delta p_{\nu}\ll p_{\nu}$),
which means that the wave packets differ not too much from, say,
plane waves. Then $\overline{\nu^{\prime}}\nu^{\prime}$  are
 very small.
We are going to study a solution when the small contribution of 
$\overline{\nu^{\prime}}\nu^{\prime}$  into the constraint
\begin{equation}
(\zeta -b)\left[sM^{4}e^{-2\alpha\phi/M_{p}}+V_{1}\right]+2V_{2}
+(\zeta +b)^{2}F_{\nu}(\zeta)\overline{\nu^{\prime}}\nu^{\prime}
=0,
\label{constraint-frw}
\end{equation}
is compensated by large value of $F_{\nu}(\zeta)$ if 
$\zeta\approx -k$:
\begin{equation}
F_{\nu}(\zeta)|_{\zeta\approx -k}=
\mu_{\nu}\frac{(h_{\nu}-k)(b-k)^{1/2}}{(\zeta +k)^{2}}+  
{\cal O}\left(\frac{1}{\zeta +k}\right).
\label{Fizeta1-k}
\end{equation}
A possible way to get up such a CLEP state might be
spreading of the wave packets during their free motion lasting
a very long (of the cosmological scale) time.
One should note however that spreading is here much more complicated 
process than in the well known quantum mechanics text books example:
 decreasing of
$\overline{\nu^{\prime}}\nu^{\prime} $ (due to the spreading) 
is related to the change of 
$\zeta$  which satisfies the nonlinear constraint equation. So we deal
in this case with nonlinear quantum mechanics.
The detailed study of  the spreading of the wave packet 
 in TMT
is a subject of considerable interest but in this paper we will 
concentrate our attention  on the consequences of the assumption
that  states with $\zeta\approx -k$  are achievable. 
We will see below that in the context of the cosmology of the
late time universe such states can provide lower energy of the universe
than the states with $\zeta$ determined  in Sec.V where there 
are no fermions at all. In other words, independently of how the
states with $\zeta\approx -k$ are realized they are energetically
 more preferable
than states without massive fermionic matter at all. 

 It is clear that it is very hard to observe  the primordial neutrinos
in the CLEP states because $\overline{\nu^{\prime}}\nu^{\prime}$
is anomalously small. Furthermore, the attempt to 
detect the fermion in the CLEP
state should be accompanied by its localization that destroy the 
condition
for the existence of such states. This is the reason to believe that 
{\bf primordial neutrinos and antineutrinos in the CLEP states might be 
good candidate for dark matter}.

\subsection{Cosmology of the late time universe with 
the cold neutrino dark matter in the CLEP state}

The study of the cosmology in the context of TMT becomes complicated
in the presence of the fermionic matter  because the averaging
procedure of the matter and gravity equations of motion includes
also the need to know the field $\zeta$ as a solution of the constraint.
As it was noticed before (see discussion after Eq.(\ref{constraint3}) )
this  generically results in the appearance of a very nonlinear 
$\overline{\Psi^{\prime}}\Psi^{\prime}$-dependence
 in all equations of motion. This makes the averaging in the cosmological
scales very unclear and practically unrealizable procedure.

Significant simplifications of the described  general situation
we observed in sections V and VIIIA were related to the fact
that in those  particular cases  the function $\zeta$ appears to be
only $\phi$-dependent or constant (or approaching constant). 

 We are going now  to apply the results of Sec.VIIIA
to see whether it is possible  to obtain a satisfactory scenario
of the late time universe where the primordial neutrinos
and antineutrinos in the
CLEP states play the role of the cold dark matter. As the first step 
we will consider here a scenario where in addition to the 
homogeneous  component of the scalar field $\phi$,  the
universe is filled also by homogeneously 
distributed\footnote{Actually one should know the
clustering properties of the proposed cold dark matter model. }
cold neutrinos and antineutrinos in the CLEP states. 
We will assume that
the cold neutrinos and antineutrinos are stable and their 
total number is
stabilized at the late time universe. Then after 
averaging over typical cosmological scales (resulting in 
the Hubble low), the constraint (\ref{constraint3})  reads
\begin{equation}
-(k+b)\left(sM^{4}e^{-2\alpha\phi/M_{p}}+V_{1}\right)
+2V_{2}+ 
(b-k)^{2}\frac{n^{(\nu)}_{0}}{a^{3}}
F_{\nu}(\zeta)|_{\zeta\approx -k}
={\cal O}\left((\zeta +k)
(sM^{4}e^{-2\alpha\phi/M_{p}}+V_{1})\right).
\label{constr-k}
\end{equation} 
where  $F_{\nu}(\zeta)|_{\zeta\approx -k}$ is defined
by Eq.(\ref{Fizeta1-k}) and $n^{(\nu)}_{0}$ is a constant
determined by the total number of the cold neutrinos
and antineutrinos.

Cosmological equations for a spatially flat FRW universe filled by
the homogeneous scalar field $\phi$ and cold neutrino dark matter 
are the following
\begin{equation}
\left(\frac{\dot{a}}{a}\right)^{2}=\frac{1}{3M_{p}^{2}}\left[\rho_{d.e.}
+\rho_{d.m.}\right]
\label{FRW-eq1}
\end{equation}
\begin{equation}
\ddot{\phi}+3\frac{\dot{a}}{a}\dot{\phi}
+\frac{2\alpha k}{(b-k)^{2}M_{p}}M^{4}e^{-2\alpha\phi/M_{p}}
+{\cal O}\left((\zeta +k)e^{-2\alpha\phi/M_{p}}\right)
=0
\label{d.e.-eq+constr}
\end{equation}
where the dark energy density $\rho_{d.e.}$ and 
dark matter energy density $\rho_{d.m.}$ are 
respectively
\begin{equation}
\rho_{d.e.}(\phi)=\frac{1}{2}\dot{\phi}^{2}+
\frac{bV_{1}-V_{2}}{(b-k)^{2}}
+\frac{b}{(b-k)^{2}}sM^{4}e^{-2\alpha\phi/M_{p}}
+{\cal O}(\zeta +k),
\label{rho-d.e.}
\end{equation}
\begin{equation}
\rho_{d.m.}=\left[\frac{\mu_{\nu}(h_{\nu}-k)}
{(\zeta +k)(b-k)^{1/2}}-
F_{\nu}(\zeta)|_{\zeta\approx -k}\right]
\frac{n^{(\nu)}_{0}}{a^{3}}
\label{rho-d.m.}
\end{equation}

The constraint (\ref{constr-k}) and
Eq.(\ref{Fizeta1-k}) show that the
corrections ${\cal O}(\zeta +k)$ behave as
\begin{equation}
{\cal O}(\zeta +k)\propto\frac{1}{a^{3/2}}
\label{cal-o}
\end{equation}
We will see that they are negligible if no fine
tuning of the parameters of the theory will be done.

We choose $s=+1$ and in addition to Eq.(\ref{bv1>v2}) assume that 
\begin{equation}
V_{1}>0, \enspace V_{2}>0 \quad and \quad
b>0, \enspace k<0, \enspace h_{\nu}<0, \enspace h_{\nu}-k<0,
 \enspace b+k<0.
\label{parameters}
\end{equation}

 In the case we work
with a positive fermion energy solution, in order to provide
positivity of the effective neutrino mass (see Eq.(\ref{muferm1}))
 in the CLEP
state, we should consider the regime where 
$\zeta =-k-\varepsilon$, $\varepsilon >0$. Then 
$F_{\nu}(\zeta)|_{\zeta\approx -k}<0$
and both of the terms in Eq.(\ref{rho-d.m.}) are positive.
The first term in Eq.(\ref{rho-d.m.}) results from the 
canonical part 
$T_{00}^{(\nu,can)}$ of the neutrino 
energy-momentum
tensor after 
making use of the equations for
 neutrino (antineutrino) field,
neglecting the terms proportional
to 3-momenta of the neutrinos  and antineutrinos and 
averaging. The second term in Eq.(\ref{rho-d.m.}) comes
from the dynamical fermionic
$\Lambda_{dyn}^{(ferm)}$ term.

The contribution of the cold neutrino dark 
matter to the pressure in the same approximation
is determined only by the dynamical fermionic
$\Lambda_{dyn}^{(ferm)}$ term and it is negative:
\begin{equation}
P_{d.m.}=
\frac{n^{(\nu)}_{0}}{a^{3}}
F_{\nu}(\zeta)|_{\zeta\approx -k}
=
-\frac{2V_{2}+|b+k|V_{1}}{(b-k)^{2}}
-\frac{|b+k|}{(b-k)^{2}}
M^{4}e^{-2\alpha\phi/M_{p}}
<0.
\label{p-d.m.}
\end{equation}   
where the constraint (\ref{constr-k}) has been used.

Further, neglecting in (\ref{rho-d.m.}) terms of the order
of $(\zeta +k)^{-1}$ as compared
to the terms of the order of $(\zeta +k)^{-2}$ 
and  using again the constraint (\ref{constr-k}) one can 
rewrite $\rho_{d.m.}$ in the form 
 \begin{equation}
\rho_{d.m.}=\frac{2V_{2}+|b+k|V_{1}}{(b-k)^{2}}
+\frac{|b+k|}{(b-k)^{2}}
M^{4}e^{-2\alpha\phi/M_{p}}
\label{rho-dm-de}
\end{equation}
typical for the dark energy sector including both 
a cosmological constant and a potential
(compare (\ref{rho-dm-de}) with (\ref{rho-d.e.})). 
The accuracy of this approximation grows as $a(t)\rightarrow\infty$.

The very important feature of {\bf the cold dark matter} realized via the 
CLEP states is that it {\bf possesses negative pressure} and in the 
context of a quintessence-like scenario (i.e. 
$\phi\rightarrow\infty$ as $a(t)\rightarrow\infty$) of the
 FRW cosmology, its equation of state approaches that
of the cosmological constant ($P_{d.m.}=-\rho_{d.m.}$) 
as $a(t)\rightarrow\infty$. In the same asymptotic regime,
the ratio $\Omega_{d.e.}/\Omega_{d.m.}$ approach 
the constant
 \begin{equation}
\frac{\Omega_{d.e.}}{\Omega_{d.m.}}\rightarrow
\frac{bV_{1}-V_{2}}{|b+k|V_{1}+2V_{2}} \quad as \quad
a(t)\rightarrow\infty 
\label{coinc-asymp}
\end{equation}
that solves (at least qualitatively) the problem
of the cosmic coincidence. Note that there is a broad range
of the parameters which allow fitting of the desirable value 
of $\frac{\Omega_{d.e.}}{\Omega_{d.m.}}\approx 2$. This value is achieved
if we assume for example that $b=\frac{2}{3}|k|$ and 
$V_{2}\ll bV_{1}$. 

The reason we regard our dark  matter model as the cold one is the 
imposed condition that neutrinos and antineutrinos
 are nonrelativistic. The latter becomes possible because the 
fermion mass in CLEP state increases as 
$m_{\nu}(\zeta)\propto (\zeta +k)^{-1}\propto a^{3/2}$.
We would like to emphasize once more that the strange enough fact, 
that  the 
CLEP-neutrino cold matter possesses negative pressure,
 is possible due to the presence of the 
noncanonical part of the fermion energy-momentum tensor,
Eq.(\ref{Tmn-noncan}), having the form of a dynamical
fermionic $\Lambda^{(ferm)}_{dyn}$ term.

The total energy density and
the total pressure of the dark sector 
(including both dark energy and dark matter) in the framework
 of the explained
 above approximations can be represented in the form
\begin{equation}
\rho^{(total)}_{dark}\equiv\rho_{d.e}+\rho_{d.m}
 =\frac{1}{2}\dot{\phi}^{2}+U_{dark}^{(total)}(\phi) ;
\quad
P^{(total)}_{dark}\equiv P_{d.e.}+P_{d.m.}
=\frac{1}{2}\dot{\phi}^{2}-U_{dark}^{(total)}(\phi),
\label{tot-rho-p-nu}
\end{equation}
where the effective potential $U_{dark}^{(total)}(\phi)$
is the sum 
\begin{equation}
U_{dark}^{(total)}(\phi)\equiv \Lambda +V_{quint}(\phi),
\label{Ueff-phi}
\end{equation}
of the cosmological constant
\begin{equation}
\Lambda =
\frac{V_{2}+|k|V_{1}}{(b-k)^{2}}
\label{Lambda-nu}
\end{equation}
and the exponential  potential
\begin{equation}
V_{quint}(\phi)=
\frac{|k|}{(b-k)^{2}}M^{4}e^{-2\alpha\phi/M_{p}}.
\label{pot-nu}
\end{equation}
This means that the evolution 
of the late time universe in the state with
$\zeta\approx -k$
proceeds as it would be in the standard field theory
 model (non-TMT) including
{\bf both the cosmological constant
 and the quintessence field $\phi$ with the exponential
potential}\footnote{Models with dark energy including
 both the cosmological constant   
 and the quintessence field has been discussed 
recently\cite{Gonzalez-Diaz}}. The emergence of the
cosmological constant guarantees an accelerated expansion
of the universe. This is why, in contrast with the standard 
quintessence scenario with exponential potential, we should not
worry here about the value of the parameter $\alpha$ in order
to provide accelerating 
expansion\footnote{The model with $V_{1}=V_{2}=0$
is shortly presented in Appendix D where the appropriate
estimations for the asymptotic ratio $\Omega_{d.e.}/\Omega_{d.m.}$
are made.}. Note that to provide the observable 
energy densities (for example, 
$\Lambda\sim\rho_{crit}$, where $\rho_{crit}$ is the present
day critical energy density) there is no need of fine tuning
of the dimensionfull parameters  $V_{1}$ and $V_{2}$ but    
instead one can assume that the dimensionless parameters
$b$, $k$ are very large. 

Summarizing what we have learned in this section and in Sec.V,
we conclude that in order to satisfy the constraint
 for  very small $|\overline{\Psi}^{\prime}\Psi^{\prime}|$
 one can proceed  only in two possible ways:
\\
i) to ignore the fermionic contribution to the constraint and solve
it for $\zeta$ in a state "absent of fermions"  or,
alternatively
\\
ii) to study the CLEP states  ($\zeta \rightarrow -k$)
where the fermionic contribution is not negligible.

It is very interesting to compare the effective potential
$V_{eff}^{(0)}(\phi)$, Eq.(\ref{Veffvac}),
predicted for the universe filled only by the homogeneous scalar field
 (for short, a state "absent of fermions"), on the one hand,
 with the effective dark sector potential $U_{dark}^{(total)}(\phi)$
 for the universe filled both by the homogeneous scalar field 
and by the homogeneous cold dark neutrino
matter (for short, "CLEP state"), on the other hand.
 The {\bf remarkable result} consists
in the fact that if $bV_{1}>V_{2}$, which is needed for positivity
of $\Lambda^{(0)}$ 
(see Eqs.(\ref{lambda-without-ferm}),(\ref{bv1>v2})), then     
\begin{equation}
V_{eff}^{(0)}(\phi)-U_{dark}^{(total)}(\phi)\equiv
\frac{\left[\frac{b+k}{2}\left(V_{1}+M^{4}e^{-2\alpha\phi /M_{p}}\right)
-V_{2}\right]^{2}}
{4(b-k)^{2}\left[b\left(V_{1}+M^{4}e^{-2\alpha\phi /M_{p}}\right)
-V_{2}\right]}>0.
\label{L-L0}
\end{equation}
This means that (for the same value of $\dot{\phi}^{2}$)
{\bf the universe in "the CLEP state" has a lower energy density
than the one in the  "absent of fermions" state}. One should 
emphasize that this result does not imply at all that $\rho_{d.m}$
is negative.

For
illustration of what kind of solutions one can expect, let us take the
{\em particular value} for the parameter $\alpha$, namely
$\alpha =\sqrt{3/8}$.
Then in the framework of the explained above approximations,
 the cosmological equations allow the following analytic solution:  
 \begin{equation}
\phi(t)=\frac{M_{p}}{2\alpha}\varphi_{0}+
\frac{M_{p}}{2\alpha}\ln(M_{p}t),
\qquad
a(t)\propto t^{1/3}e^{\lambda t},
\label{a-sol-nu}
\end{equation}
where
\begin{equation}
\lambda =\frac{1}{M_{p}}\sqrt{\frac{\Lambda}{3}},
\qquad e^{-\varphi_{0}}=
\frac{2(b-k)^{2}M_{p}^{2}}{\sqrt{3}|k|M^{4}}\sqrt{\Lambda}.
\label{phi-0}
\end{equation}
and $\Lambda$ is determined by Eq.(\ref{Lambda-nu}).
The mass of the neutrino in such CLEP state increases
exponentially in time and its $\phi$ dependence is
double-exponential:
\begin{equation}
m_{\nu}|_{CLEP}\sim (\zeta +k)^{-1}\sim a^{3/2}(t)\sim
t^{1/2}e^{\frac{3}{2}\lambda t}\sim 
\exp\left[\frac{3\lambda e^{-\varphi_{0}}}{2M_{p}}
\exp\left(\frac{2\alpha}{M_{p}}\phi \right)\right].
\label{m-t-phi}
\end{equation}
In  this particular case, the time dependence of 
the ratio $\Omega_{d.e.}/\Omega_{d.m.}$ reads
 \begin{equation}
\frac{\Omega_{d.e.}}{\Omega_{d.m.}}\approx
\frac{bV_{1}-V_{2}+\frac{2}{3|k|}b(b+|k|)(V_{2}+|k|V_{1})^{1/2}
\cdot\frac{M_{p}}{t}}
{|b+k|V_{1}+2V_{2}+\frac{2}{3|k|}(k^{2}-b^{2})(V_{2}+|k|V_{1})^{1/2}
\cdot\frac{M_{p}}{t}},
\label{coinc-3/8}   
\end{equation}
where we have neglected corrections of the order $t^{-2}$ coming
from the kinetic term of the scalar field $\phi$.

\section{Discussion and Conclusion}
\subsection{Summary and some possible directions of  possible developments}

In this paper we have continued our study of the two measures theory,
including fermions. Except for three special physical requirements,
the theory starts from the most 
general action which possesses global scale symmetry (\ref{stferm})
and contains one primordial fermion field for each type of leptons and quarks,
i.e. does not incorporate fermion generations. It turns out that each of the 
primordial fermion fields can be in different states depending on the
fermion density. The states of the primordial fermion are controlled
by the constraint (\ref{constraint3}) which appears as the consistency
 condition of the classical equations of motion.

 In the regime of the "high fermion energy density"
which is always realized in the normal particle, nuclear and atomic physics
 experiments, 
the theory predicts splitting of each type of the leptons and quarks
exactly into  three mass eigenstates.  
So, TMT appears to be able to provide the resolution of the fermion
families problem.  Under the same "high fermion energy density" conditions, the 
 fermion-dilaton coupling suffers an extremely high suppression
that means that the fermion
families problem is resolved simultaneously with  the fifth force problem.

If the fermion energy density is of
the order of dark energy density (which we call the regime of low 
fermion energy 
density) it has been shown in this paper that the primordial  fermion does 
not split into generations and also it can be in a CLEP state where it 
participates directly in the cosmological expansion by means of changing 
its own parameters (like a mass).
  
The explicit solution of equations of motion was presented where
uniformly distributed nonrelativistic neutrinos and antineutrinos in CLEP
state appear as a good model for cold dark matter scaling like a dark
energy. The total equation of state provides accelerated expansion of the
universe.
Another attractive feature of this scenario is that the total energy density
of such universe is less than that in the case of the universe free of the
fermions at all.   

Together with the constraint (\ref{constraint3}), the key role both in 
the "high fermion energy density" (regular fermion
families and resolution of the fifth force problem)
and in the low fermion energy density (cosmo-particle solution) belongs to 
the dynamical fermionic $\Lambda^{(ferm)}_{dyn}$ term, Eq(\ref{Tmn-noncan}).
 The constraint
dictates that if $\zeta$ is $\overline{\Psi^{\prime}}\Psi^{\prime}$   
independent (which is true in all problems studied in the present paper
except for the case of the third fermion family solution),
then the value of $\Lambda^{(ferm)}_{dyn}$ must be
of the order of the
dark energy density. If the canonical fermion energy density is much larger
than $\Lambda^{(ferm)}_{dyn}$ then we deal with the "high fermion energy 
density" case where the primordial fermions are in the mass eigenstates
of the first two families of the regular fermion matter. 
However, if the canonical fermion energy
density is much less 
than $\Lambda^{(ferm)}_{dyn}$ which we call the low fermion energy density regime,
then besides the homogeneous cosmological solutions studied in this paper,
there are many inhomogeneous regimes of great interest from the
astrophysical point of view. The  important feature of the theory is that
the constraint is a local algebraic equation and therefore it dictates
not only the same scaling in time  for fermion and dark energy density
at the late time universe     
but also the space fluctuations of the fermion and
dilaton energy densities turn out to be strongly correlated in
 the low fermion energy density regime. Besides, in this regime there is
no suppression of the fermion-dilaton coupling and
therefore the fermion (in our scenario, neutrino) dark matter should be 
self-interacting via scalar force.   Taking into account also that the
(primordial) fermionic matter in this regime has negative 
pressure\footnote{The negative pressure is needed apparently for
explaining the flat rotation curves (see for example \cite{Bharadwaj})}, 
one should expect the emergence of new types of mechanisms for clumping
of dark matter and structure
formation. These intriguing questions should be the subject of future
investigations.

\subsection{Similarities with ideas and/or results discussed by 
other authors}

It is interesting to notice that in a certain sense the theory being
studied here (where nonlinear effects are found at very low energy  
densities) resembles the nonlinear theory at very low
accelerations (MOND)\cite{Milgrom}. The details are quite different,
but fact remains that nonlinear effects make the difference (i.e.   
the appearance of dark matter or dark matter effects) in both cases 
just in the weak limit.

The theory discussed here displays some of the features associated
to the Chaplygin gas models\cite{Chaplygin}. In particular it is known
that such "gas" behaves like normal matter (dust) at high densities
and like vacuum energy at low densities. Also this takes place 
in our TMT model where normal particle states emerge at high
fermion densities and for low fermion densities we have shown
the universe prefers to be in a state where the contribution of
fermionic matter is asymptotically dominated by the dynamical
fermion $\Lambda$ term.

In this paper we have found that particles are able to react to 
the cosmological evolution and change their masses, according 
to the constraint equation. This constraint in turn implies a
cosmic coincidence. All this is realized in a way that very much 
resembles the steady state model\cite{Bondi}, discarded as a 
description of the whole history of the universe, but which in
our case it seems to reappear as a good approximation for
the asymptotic behavior of the late universe. Here the 
"continuous matter creation"\cite{Bondi} is replaced not only by the 
variable particle masses (similar to VAMPS models) but rather by a
much more important effect related with  the role of 
$\Lambda^{(ferm)}_{dyn}$ term which is originated by fermionic
(neutrino) matter and behaves as a cosmological "constant".
In the theory studied here, the coincidence of dark energy and
dark matter densities can be achieved in a way we consider  
where the asymptotic values of the dark energy and dark
matter densities are constants.
 This indeed strengthens  the
similarity between our asymptotic state and the steady state 
models. 

We should point out the interesting convergence between some of our
results and those of the authors of Ref.\cite{Fardon}, where 
also the possibility
of mass varying neutrinos was considered. Although their starting 
point is rather different, many of their results actually 
are qualitatively similar to ours: (i) the neutrino can
behave as a negative pressure fluid; (ii) the neutrino mass
is correlated to the neutrino density, etc..

Finally, we find that both in the case of high fermion density and
for low fermion density, the presence of the fermions changes 
drastically the vacuum in which the fermions live. The effects
here are rather dramatic: in the high density it is responsible
for the three generations splitting of the primordial fermions
and at low densities - for the cosmic coincidence between
dark energy and dark matter densities. For more conventional 
type of field theories, the question of whether a fermion can 
change the properties of the vacuum in which it lives,
was studied in Ref.\cite{MWZ} and found that this can be 
indeed the case. The authors of Ref.\cite{MWZ} have 
demonstrated this by
obtaining a substantially different fermion mass 
in the self-consistent approach as
  compared to the one 
obtained if one had considered such a fermion as a simple
"test particle". 

\section{Acknowledgements}

We thank A. Davidson for very useful conversations and
for the idea of a possible connection of the measure $\Phi$
with brane theory in extra dimensions and for showing us
 an explicit calculation concerning this idea. We also thank 
L. Laperashvili for suggesting us the possible brane origin
of the measure $\Phi$ and for informing us on the recent 
developments concerning $6t+6\bar{t}$ condensates.
 It is a pleasure to acknowledge
extensive discussions with J. Bekenstein and M. Milgrom
on different aspects of the theory described here. One 
of us (E.G.) want to thank H. Nielsen and E. Spallucci for 
conversations and the University of Trieste for hospitality.
  
\bigskip

\appendix

\section{Possible origin of TMT from low energy limit
of the brane-world scenario}
Continuing  discussion of the general structure of the action
in TMT one of course may ask: why 3-brane moving only
in $3+1$ dimensional
embedding space-time? This can be obtained also starting from higher
dimensional "brane-world scenarios".  Indeed, let us consider for example
 a 3-brane evolving in an embedding 5-dimensional space-time with
\begin{equation}   
ds^{2}=G_{AB}dx^{A}dx^{B}=f(y)\hat{g}_{\mu\nu}(x^{\alpha})dx^{\mu}dx^{\nu}+
\gamma^{2}(y)dy^{2}, \quad  A,B=0,1,...4; \quad \mu,\nu =0,1,2,3.
\label{interval}
\end{equation}
Assuming that it is possible to ignore
the motion of the brane in the extra dimension, i.e.
studying the brane with a fixed position in extra dimension, $y=const$,
one can repeat the arguments of Sec.II (starting with the Nambu-Goto action)
 where one needs to use only
four functions $\varphi^{a}(x^{\mu})$ which together with the fifth
($x^{\mu}$ independent) component along the axis $y$ constitute the
 5-vector describing the embedding of our brane; 
$L_{1}$ is again (as in Sec.II) the
$\varphi^{a}$ independent
Lagrangian of the gravity\footnote{Notice that connection coefficients of
our 3-brane are those where indexes run only from zero to three and 
these components do not suffer from discontinuities across the 
brane\cite{Israel}} and matter on the brane. 
 As a result we obtain
exactly the same effective action as in Eq.(\ref{S-red-NG-L1}) which
describes a brane
moving in the hypersurface $y=const$ of the 5-dimensional space-time.

The brane theory action contains also a piece
coming from the bulk dynamics. We will assume that gravity and
matter exist
also in the bulk where their action can be written in the form
\begin{equation}
S_{bulk}=\int\sqrt{-\hat{g}}f^{2}(y)\gamma(y)
L_{bulk}d^{4}xdy
\label{s-bulk}
\end{equation}
where $\hat{g}\equiv det(\hat{g}_{\mu\nu})$ and $L_{bulk}$ is the Lagrangian of
the gravity and matter in
the bulk. We are not interested in the dynamics in the hole bulk   
but rather in the effect of the bulk on the 4-dimensional dynamics. For this
purpose one can integrate out the perpendicular coordinate $y$
in the action (\ref{s-bulk}). One can think of this integration
 in a spirit of  the procedure known as
averaging (for the case of compact extra dimensions see for example 
Ref.\cite{S-Weinberg}).  We do not perform
this integration explicitly\footnote{The correspondent calculations
must take into account the discontinuity constraints.} here but we expect 
that the resulting averaged 
contribution
of the bulk dynamics to the 4-dimensional action one can write down 
in the form
\begin{equation}
S_{2}=\int\sqrt{-\hat{g}}
L_{2}(\hat{g}_{\mu\nu}, connection, matter fields)d^{4}x
\label{s-brane-from bulk}
\end{equation}
that we would like to use as the 
 second term of the postulated in Eq.(\ref{S})
general form of the TMT action in four dimensions. 

Notice however that the
geometrical objects in these two actions may not be identical.
The simplest assumption would be of course to take 
$\hat{g}_{\mu\nu}\equiv g_{\mu\nu}$ (and also coinciding connections).
One may allow for the case $\hat{g}_{\mu\nu}\neq g_{\mu\nu}$
nevertheless. In fact, brane theory allows naturally bimetric
theories\cite{Damour-Kogan} even if one starts with a single 
bulk metric. In our case this can be due to the fact that the
metric at the brane and its average value may be different.
The bimetric theories\cite{Salam} give in any case only one massless
linear combination of the two metrics which one can identify as
long distance gravity. Connections  at the brane 
and its 
average value may be different as well. But we expect
this difference to be small due to the continuity of the 
relevant connection coefficients (see footnote 11).  
Therefore performing the integration over
extra dimension $y$ we are left with just one independent 
connection which is important for TMT where the first order
formalism is supposed to be one of the basic principles.
All these arguments are now being developed in details
and will be reported elsewhere.

\section{Equations of motion and constraint in
the original frame}

Equation (\ref{varphi}) corresponding to the model 
(\ref{totaction}) reads:
\begin{equation}
\left[-\frac{1}{\kappa}R(\omega, e)+ 
\frac{1}{2}g^{\mu\nu}\phi_{,\mu}\phi_{,\nu}\right]e^{\alpha\phi /M_{p}}
-V_{1}e^{2\alpha\phi /M_{p}} +L_{fk} e^{\alpha\phi /M_{p}}
-\sum_{i=\nu ,E}\mu_{i}
\overline{\Psi}_{i}\Psi_{i} e^{\frac{3}{2}\alpha\phi /M_{p}} 
=sM^{4},
\label{app1}
\end{equation}
where
\begin{equation}
L_{fk}=\frac{i}{2}
\left[\overline{\nu}
\left(\gamma^{a}e_{a}^{\mu}\overrightarrow{\nabla}_{\mu}
-\overleftarrow{\nabla}_{\mu}\gamma^{a}e_{a}^{\mu}\right)\nu
+\overline{E}
\left(\gamma^{a}e_{a}^{\mu}\overrightarrow{\nabla}^{(E)}_{\mu}
-\overleftarrow{\nabla}^{(E)}_{\mu}\gamma^{a}e_{a}^{\mu}\right)E
\right].
\label{app2}
\end{equation}

Variation of the action (\ref{totaction}) with respect to 
$e^{a\mu}$ yields
\begin{eqnarray}
&&(\zeta +b)\left[-\frac{2}{\kappa}R_{a,\mu}(\omega ,e)
+e^{\beta}_{a}\phi_{,\mu}\phi_{,\beta}\right]
\nonumber\\
&+& g_{\mu\beta}e^{\beta}_{a}\left[\frac{b}{\kappa}R(\omega ,e)
-\frac{b}{2}g^{\mu\nu}\phi_{,\mu}\phi_{,\nu}
+V_{2}e^{\alpha\phi /M_{p}}-kL_{fk}
+(h_{\nu}\mu_{\nu}\overline{\nu}\nu
+h_{E}\mu_{E}\overline{E}E)e^{\frac{1}{2}\alpha\phi /M_{p}}
\right]
\nonumber\\
&+& (\zeta +b)\frac{i}{2}
\left[\overline{\nu}
\left(\gamma^{a}\overrightarrow{\nabla}_{\mu}
-\overleftarrow{\nabla}_{\mu}\gamma^{a}\right)\nu 
+\overline{E}
\left(\gamma^{a}\overrightarrow{\nabla}^{(E)}_{\mu}
-\overleftarrow{\nabla}^{(E)}_{\mu}\gamma^{a}\right)E
\right]
+\frac{1}{\sqrt{-g}}\frac{\partial\sqrt{-g}L_{em}}
{\partial e^{a,\mu}}e^{-\alpha\phi /M_{p}}=0
\label{app3}
\end{eqnarray}
where 
$L_{em}=
-\frac{1}{4}g^{\alpha\beta}g^{\mu\nu}F_{\alpha\mu}F_{\beta\nu}$.

Contraction of Eq.(\ref{app3}) with $e^{a,\mu}$ gives
\begin{equation}
2(\zeta -b)\left[-\frac{1}{\kappa}R(\omega, e)+
\frac{1}{2}g^{\mu\nu}\phi_{,\mu}\phi_{,\nu}\right]
+4V_{2}e^{\alpha\phi /M_{p}} +(\zeta -3k)L_{fk} 
+4(h_{\nu}\mu_{\nu}\overline{\nu}\nu
+h_{E}\mu_{E}\overline{E}E)e^{\frac{1}{2}\alpha\phi /M_{p}}
=0,
\label{app5}
\end{equation}
where identity $e^{a\mu}(\partial
\sqrt{-g}L_{em}/\partial e^{a\mu})\equiv 0$ has been used.

Excluding $R(\omega, e)$ from Eqs.(\ref{app2}) and (\ref{app5})
we obtain consistency condition of these two equations:
\begin{eqnarray}
&& 2(\zeta -b)\left(sM^{4}e^{-\alpha\phi /M_{p}}+V_{1}e^{\alpha\phi /M_{p}}\right)
+4V_{2}e^{\alpha\phi /M_{p}}
-(\zeta-2b+3k)L_{fk}
\nonumber\\
&+& 2[(\zeta-b+2h_{\nu})\mu_{\nu}\overline{\nu}\nu
+(\zeta-b+2h_{E})\mu_{E}\overline{E}E)]e^{\frac{1}{2}\alpha\phi /M_{p}}
=0,
\label{app6} 
\end{eqnarray}

Varying $\overline{\nu}$ and $\nu$ we get:
\begin{eqnarray}
&& (\zeta +k)\frac{i}{2} \left[\gamma^{a}e_{a}^{\beta} (\overrightarrow{\partial}_{\beta}+
\frac{1}{2}\omega_{\beta}^{cd}\sigma_{cd})+
\frac{1}{2}\omega_{\beta}^{cd}\sigma_{cd}\gamma^{a}e_{a}^{\beta}\right]\nu
\nonumber\\
&+&\frac{i}{2\sqrt{-g}}e^{-\alpha\phi /M_{p}}
\partial_{\beta} \left[\sqrt{-g}(\zeta +k)e^{\alpha\phi /M_{p}}\gamma^{a}e_{a}^{\beta}\nu\right]
-(\zeta +h_{\nu})\mu_{\nu}\nu e^{\frac{1}{2}\alpha\phi /M_{p}}
=0
\label{app7}
\end{eqnarray}
\begin{eqnarray}
&& (\zeta +k)\frac{i}{2} \left[-\overline{\nu}(\overleftarrow{\partial}_{\beta}-
\frac{1}{2}\omega_{\beta}^{cd}\sigma_{cd})\gamma^{a}e_{a}^{\alpha} 
+\overline{\nu}\gamma^{a}e_{a}^{\beta}\frac{1}{2}\omega_{\beta}^{cd}\sigma_{cd}\right]
\nonumber\\
&-&\frac{i}{2\sqrt{-g}}e^{-\alpha\phi /M_{p}}
\partial_{\beta} \left[\sqrt{-g}(\zeta +k)e^{\alpha\phi /M_{p}}
\overline{\nu}\gamma^{a}e_{a}^{\beta}\right]
-(\zeta +h_{\nu})\mu_{\nu}\overline{\nu}e^{\frac{1}{2}\alpha\phi /M_{p}}
=0
\label{app8}
\end{eqnarray}
and similarly for $E$ and $\overline{E}$. Multiplying the last two
equations by $\overline{\nu}$ and ${\nu}$ respectively (and
similarly for $E$ and $\overline{E}$) we obtain
\begin{equation}
L_{fk}=\frac{1}{(\zeta +k)}
[(\zeta +h_{\nu})\mu_{\nu}\overline{\nu}\nu
+(\zeta +h_{E})\mu_{E}\overline{E}E]e^{\frac{1}{2}\alpha\phi /M_{p}}
\label{app9}
\end{equation}

Inserting $L_{fk}$ into the consistency condition, Eq.(\ref{app6}),
we get {\em the constraint (in the original frame)} as the following
algebraic equation
\begin{equation}
(\zeta -b)\left(sM^{4}e^{-2\alpha\phi /M_{p}}+V_{1}\right)
+2V_{2}
+\frac{e^{\frac{1}{2}\alpha\phi /M_{p}}}{2(\zeta +k)}\sum_{i=\nu, E} 
f_{i}(\zeta)\mu_{i}\overline{\Psi}_{i}\Psi_{i}=0.
\label{app10}
\end{equation}
where
\begin{equation}
f_{i}(\zeta)\equiv\zeta^{2}
+(3h_{i}-k)\zeta +2b(h_{i}-k)+kh_{i}, \quad i=\nu, E
\label{app11}
\end{equation}

Contracting Eq.(\ref{app3}) with factor $e^{a}_{\nu}$
and using  Eq.(\ref{app9})
we get the gravitational equations in the original frame
\begin{eqnarray}
 \frac{2}{\kappa}R_{\mu\nu}(\omega, e)&=&\phi_{,\mu}\phi_{,\nu}
-g_{\mu\nu}\frac{1}{\zeta +b}\left(bsM^{4}e^{-\alpha\phi /M_{p}}
      +(bV_{1}-V_{2})e^{\alpha\phi /M_{p}}\right)  
\nonumber\\
&+& \frac{\zeta +k}{\zeta +b}
\frac{i}{2}
\sum_{i=\nu,E}\overline{\Psi}_{i}
\left(\gamma^{a}e_{a \nu}\overrightarrow{\nabla}_{\mu}
-\overleftarrow{\nabla}_{\mu}\gamma^{a}e_{a \nu}\right)\Psi_{i} 
+g_{\mu\nu}\frac{1}{\zeta +k}e^{\frac{1}{2}\alpha\phi /M_{p}}
\sum_{i=\nu, E}(h_{i}-k)\mu_{i}\overline{\Psi}_{i}\Psi_{i}
\label{app12}
\end{eqnarray}
where for short we have omitted term with $F_{\mu\nu}$
since it is clear that due to the conformal invariance of 
$L_{em}$ its contribution 
to the energy-momentum tensor will be  canonical in the Einstein 
frame.

The scalar field $\phi$ equation of motion in the original frame
can be written in the form
\begin{eqnarray}
&& \frac{1}{\sqrt{-g}}\partial_{\mu}\left[e^{\alpha\phi /M_{p}}
(\zeta +b)\sqrt{-g}g^{\mu\nu}\partial_{\nu}\phi\right]
\nonumber\\
&-& \frac{\alpha}{M_{p}}\left[(\zeta +b)sM^{4}+
\left((b-\zeta)V_{1}-2V_{2}\right)e^{2\alpha\phi /M_{p}}
-\frac{e^{\frac{3}{2}\alpha\phi /M_{p}}}{2(\zeta +k)}
\sum_{i=\nu, E}f_{i}(\zeta)\mu_{i}\overline{\Psi}_{i}\Psi_{i}
\right]
\label{app13}  
\end{eqnarray}
where Eqs.(\ref{app1}), (\ref{app9}) and the notation
 (\ref{app11}) have been used.

\bigskip
\section{Connection in the original frame}
\bigskip

We present here what is the dependence of the spin connection
$\omega_{\mu}^{ab}$ on $e^{a}_{\mu}$, $\zeta$, $\Psi$ and $\overline{\Psi}$.
Varying the action (\ref{totaction}) with respect to
$\omega_{\mu}^{ab}$ and making use that
\begin{equation}
R(V,\omega)\equiv
-\frac{1}{4\sqrt{-g}}\varepsilon^{\mu\nu\alpha\beta}\varepsilon_{abcd}
e^{c}_{\alpha}e^{d}_{\beta}R_{\mu\nu}^{ab}(\omega)
\label{C1}
 \end{equation}  
we obtain
\begin{equation}
\varepsilon^{\mu\nu\alpha\beta}\varepsilon_{abcd}[(\zeta +b)
e^{c}_{\alpha}D_{\nu}e^{d}_{\beta}
+\frac{1}{2}e^{c}_{\alpha}v^{d}_{\beta}\zeta,_{\nu}]+
\frac{\kappa}{4}\sqrt{-g}\frac{\zeta +k}{\zeta +b}e^{c\mu}\varepsilon_{abcd}\overline{\Psi}
\gamma^{5}\gamma^{d}\Psi=0,   
\label{C2} 
 \end{equation}
where
\begin{equation}
D_{\nu}e_{a\beta}\equiv\partial_{\nu}e_{a\beta}
+\omega_{\nu a}^{d}e_{d\beta}
\label{C3}
 \end{equation}
The solution of Eq. (\ref{C2}) is represented in the form
\begin{equation}
\omega_{\mu}^{ab}=\omega_{\mu}^{ab}(e)   +
K_{\mu}^{ab}(e,\overline{\Psi},\Psi)
+ K_{\mu}^{ab}(\zeta)
\label{C4}
 \end{equation}
where
\begin{equation}
\omega_{\mu}^{ab}(e)=e_{\alpha}^{a}e^{b\nu}\{ ^{\alpha}_{\mu\nu}\}-
e^{b\nu}\partial_{\mu}e_{\nu}^{a}
\label{C5}  
 \end{equation}
is the Riemannian part of the spin-connection,
\begin{equation}
K_{\mu}^{ab}(e,\overline{\Psi},\Psi)=
\frac{\kappa}{8}\frac{\zeta +k}{\zeta +b}
\eta_{cn}e_{d\mu}\varepsilon^{abcd}\overline{\Psi}
\gamma^{5}\gamma^{n}\Psi
\label{C7}
 \end{equation}
is the fermionic contribution\cite{Gasperini} 
and
\begin{equation}
K_{\mu}^{ab}(\zeta)=\frac{1}{2(\zeta +b)}\zeta_{,\alpha}(e_{\mu}^{a}e^{b\alpha}-
e_{\mu}^{b}e^{a\alpha})
\label{C6}
 \end{equation}
is {\bf the non-Riemannian part of the spin-connection originated by
specific features of TMT}.

The term $K_{\mu}^{ab}(e,\overline{\Psi},\Psi)$ is
responsible for a correction
to the Einstein equations (in the Einstein frame)
having the form of the so-called spin-spin contact
interaction\cite{Gasperini} and proportional
to the square of the Newton constant. 
Notice also that in the context of homogeneous cosmology this contribution
 is absent at all.
If some inhomogeneities present, this contribution has
additional suppression in the regime $\zeta\approx -k$.

\bigskip
\section{A model without explicit potentials}
\bigskip

In the recent paper\cite{GK4} we have studied the model
similar to one of the present paper but without explicit
potentials in the action ($V_{1}=V_{2}=0$ in Eq.(\ref{totaction}).
In this case the effective exponential potential is 
generated by the spontaneous symmetry breaking discussed
in Sections II and IV of the present paper. The aim of this 
Appendix is to check whether it is possible, in the context
of the model described by the action (\ref{totaction})
where now we take $V_{1}=V_{2}=0$,  to provide conditions
both for an accelerated expansion and for 
a resolution of the cosmic coincidence problem in 
a way similar to what we have seen Sec.VIII, that is in
the framework of a scenario
with the cold neutrino dark matter in CLEP state.

Equations of motion and constraint of the model have the form
of equations of Sec.IV where now $V_{1}=V_{2}=0$. For short
we will not write down the new equations here. Instead, 
in this Appendix we
will refer to equations of motion and constraint of Sec.IV
 keeping in mind that now we have $V_{1}=V_{2}=0$. 

Let us start from a simple notion that in the case
of the absence of fermions, i.e. when we deal with a spatially flat
 FRW universe filled only by
the homogeneous scalar field $\phi$,
the constraint (\ref{constraint3}) yields $\zeta =b$.
Therefore we have in this case a standard quintessence model with
the exponential potential
\begin{equation}
V_{eff}^{(0)}(\phi)=\frac{M^{4}}{4b}e^{-2\alpha\phi/M_{p}}.
\label{quint-stand}
\end{equation}

Turn now to a spatially flat FRW universe filled by
the homogeneous quintessence field $\phi$ and cold neutrino dark 
matter in the CLEP state.
The respective cosmological equations read
(we choose the same restrictions on the dimensionless
parameters as in Eq.(\ref{parameters}))
\begin{equation}
\left(\frac{\dot{a}}{a}\right)^{2}=\frac{1}{3M_{p}^{2}}[\rho_{q}
+\rho_{d.m.}] 
\label{FRW-eq}
\end{equation}
\begin{equation}
\ddot{\phi}+3\frac{\dot{a}}{a}\dot{\phi}
-\frac{2\alpha |k|}{(b-k)^{2}M_{p}}M^{4}e^{-2\alpha\phi/M_{p}}
+{\cal O}\left((\zeta +k)e^{-2\alpha\phi/M_{p}}\right)
=0,
\label{quint-eq}
\end{equation}
where 
the quintessence energy density is 
\begin{equation}
\rho_{q}(\phi)=\frac{1}{2}\dot{\phi}^{2}+\frac{b}{(b-k)^{2}}
M^{4}e^{-2\alpha\phi/M_{p}}
+{\cal O}\left((\zeta +k)e^{-2\alpha\phi/M_{p}}\right)
\label{rho-q}
\end{equation}
and the dark matter energy density has the same form as in
Eq.(\ref{rho-d.m.}) with the function 
$F_{\nu}(\zeta)|_{\zeta\approx -k}$ determined by 
Eq.(\ref{Fizeta1-k}). 

The constraint that reads now
\begin{equation}
-(k+b)M^{4}e^{-2\alpha\phi/M_{p}}+  
(b-k)^{2}\frac{n^{(\nu)}_{0}}{a^{3}}
F_{\nu}(\zeta)|_{\zeta\approx -k}
={\cal O}\left((\zeta +k)M^{4}e^{-2\alpha\phi/M_{p}}\right)
\label{constr-k-00}
\end{equation}
allows to represent the dark matter energy density as the 
function of the quintessence field:
\begin{equation}
\rho_{d.m.}=\frac{(b+k)}{(b-k)^{3}}
M^{4}(\zeta +k) e^{-2\alpha\phi/M_{p}}
-\frac{b+k}{(b-k)^{2}}M^{4}e^{-2\alpha\phi/M_{p}}
+{\cal O}\left(\frac{1}{(\zeta +k)a^{3}}\right).
\label{rho-d-00}
\end{equation}
It follows from the constraint (\ref{constr-k-00})
 that behaviour both of the first term
in (\ref{rho-d-00}) and of the corrections 
${\cal O}\left(\frac{1}{(\zeta +k)a^{3}}\right)$
 are of the form
\begin{equation}
(\zeta +k)e^{-2\alpha\phi/M_{p}}\propto 
{\cal O}\left(\frac{1}{(\zeta +k)a^{3}}\right)
\propto \frac{e^{-\alpha\phi/M_{p}}}{a^{3/2}}.
\label{rho-d-00-corr}
\end{equation}
The same order of corrections have appeared both
in Eq.(\ref{FRW-eq}) and in Eq.(\ref{quint-eq}).
As we will see below, if {\em one to demand} that solutions of
the cosmological equations describe {\em an accelerated expansion}
of the universe, this is sufficient to provide that
all such corrections decrease
more rapidly than the rest of the terms. This is why studying the
late time (accelerated) universe one can regard all these
corrections as negligible.

Ignoring the above
mentioned corrections, we have a situation where the dark
matter contribution to the energy-momentum tensor behaves as
an additional exponential potential and the dark matter
equation of state approaches that of the cosmological
constant ($P_{d.m.}=-\rho_{d.m.}$) as $a(t)\rightarrow\infty$.
As a result of this,
 the total energy density and total pressure are
in this approximation:
\begin{equation}
\rho_{tot}\equiv \rho_{q}+\rho_{d.m.}=
\frac{1}{2}\dot{\phi}^{2}+\frac{|k|}{(b-k)^{2}}
M^{4}e^{-2\alpha\phi/M_{p}}
\label{rho-tot-00}
\end{equation}  
 \begin{equation}
P_{tot}\equiv P_{q}+P_{d.m.}=
\frac{1}{2}\dot{\phi}^{2}-\frac{|k|}{(b-k)^{2}}
M^{4}e^{-2\alpha\phi/M_{p}}
\label{p-tot-00}
\end{equation}
This means that in the late time universe
the cosmological equations
(\ref{FRW-eq}), (\ref{quint-eq}) are reduced to the standard ones
of the quintessential cosmology with the exponential potential   
$\frac{|k|}{(b-k)^{2}}M^{4}e^{-2\alpha\phi/M_{p}}$.
It is easy to show that similar to what we have seen 
in  the model of Sec.VIIIB, Eq.(\ref{L-L0}),
{\em the universe in "the CLEP state" has a lower energy density
than the one in the  "absent of fermions" state}.

The correspondent cosmological solution is the following:
\begin{equation}
\phi(t)=\frac{M_{p}}{2\alpha}\varphi_{0}+
\frac{M_{p}}{\alpha}\ln(M_{p}t),
\qquad
a(t)\propto (M_{p}t)^{1/2\alpha^{2}}, \qquad
e^{-\phi_{0}}=\frac{(3-2\alpha^{2})(b-k)^{2}M_{p}^{4})}
{4\alpha^{4}|k|M^{4}}.  
\label{quint-sol-00}
\end{equation}

According to this solution, the universe expands with acceleration
if $\alpha <\frac{1}{\sqrt{2}}$.
In such a case the corrections we
ignored solving the above cosmological equations, behave in time
as
\begin{equation}
\frac{1}{(\zeta +k)a^{3}}\propto
(\zeta +k)e^{-2\alpha\phi/M_{p}}\propto \frac
{e^{-\alpha\phi/M_{p}}}{a^{3/2}}\propto
t^{-(1+3/4\alpha^{2})}<\frac{1}{t^{5/2}}.
\label{rho-d1-first}
\end{equation}  

The time dependence  of
the quintessence and dark matter energy densities 
corresponding to the solution (\ref{quint-sol-00}) are respectively:
\begin{equation}
\rho_{q}= \frac{1}{4\alpha^{4}|k|}
[2(|k|-b)\alpha^{2}+3b]
\cdot\frac{M_{p}^{2}}{t^{2}},
\label{rho-q-t} 
\end{equation}
\begin{equation}
\rho_{d.m.}= \frac{1}{4\alpha^{4}|k|}
(|k|-b)(3-2\alpha^{2})
\cdot
\frac{M_{p}^{2}}{t^{2}}
\label{rho-d-t}
\end{equation}
Then for the ratio
\begin{equation}
\frac{\Omega_{q}}{\Omega_{d.m.}}=
\frac{2(|k|-b)\alpha^{2}+3b}
{(|k|-b)(3-2\alpha^{2})}
\label{coinc-00}   
\end{equation}
in the case of accelerated expansion ($\alpha^{2}<1/2$)
we obtain
\begin{equation}
\frac{\Omega_{q}}{\Omega_{d.m.}}<
\frac{|k|+2b}
{2(|k|-b)}.
\label{coinc-00-estim}   
\end{equation}
So, we conclude that the model without explicit
potentials provides conditions  for a cosmic coincidence in
the scenario of the accelerated expansion of the
late time universe.

The time dependence of the mass of the neutrino in 
CLEP state is
\begin{equation}
m_{\nu}|_{CLEP}\sim t^{-1+3/4\alpha^{2}}
\label{m-t}
\end{equation}
and in the scenario of an accelerating expansion it
increases in time faster than $t^{1/2}$.

\end{document}